\documentclass[aps,showpacs,preprint,preprintnumbers,superscriptaddress,nofootinbib]{revtex4}

 \usepackage{graphicx,amsmath}
 \usepackage{enumerate}
\usepackage{multirow}

\usepackage{hyperref}

\def\Journal#1#2#3#4{{#1} {{\bf #2},} {#4} {(#3)}}

\def\NP{{Nucl. Phys.} }

\def\PLB{{Phys. Lett.}  B}
\def\PL{{Phys. Lett.}}

\def\PRP{{ Phys. Rep.}}
\def\PRL{ Phys. Rev. Lett.}

\def\PRD{{Phys. Rev.} D}
\def\PRC{{Phys. Rev.} C}

\def\ZPC{{Z. Phys. C}}

\def\EPJC{{Eur. Phys. J.} C}

\def\MPLA{{Mod. Phys. Lett.} A}

\def\JHEP{{J. High Energy Phys.}}
\def\JPG{{J. Phys. G.}}
\def\SJNP{Sov. J. Nucl. Phys.}

\def\ra{\rightarrow}

\def\be{\begin{equation}}
\def\ee{\end{equation}}
\def\bea{\begin{eqnarray}}
\def\eea{\end{eqnarray}}

\def\qbar{{\bar q}}

\newcommand{\mbf}[1]{\mathbf{#1}}

\usepackage{relsize}
\def\babar{\mbox{\slshape B\kern-0.1em{\smaller A}\kern-0.1em
    B\kern-0.1em{\smaller A\kern-0.2em R}}}

\def\NP{{ Nucl. Phys.}}

\begin{document}

\preprint{SLAC-PUB-14374}

\title{ Evolved QCD predictions for the meson-photon transition form factors}

\author{Stanley J. Brodsky}

\affiliation{SLAC National Accelerator Laboratory, Stanford University, California 94309, USA}

\author{Fu-Guang Cao}

\affiliation{Institute of Fundamental Sciences, Massey University, Private Bag 11 222, \\ Palmerston North, New Zealand}
\affiliation{College of Nuclear Science and Technology, Beijing Normal University, Beijing 100875, P. R. China}
\thanks{On leave from Massey University, New Zealand.}

\author{Guy F. de T{\'e}ramond}
\affiliation{Universidad de Costa Rica, San Jos\'e, Costa Rica}

\date{\today}

\begin{abstract}
The QCD evolution of  the pion distribution amplitude (DA) $\phi_\pi(x,Q^2)$ is computed for several commonly used models.
Our analysis includes the nonperturbative form predicted by light-front holographic QCD,
thus combining the nonperturbative bound state dynamics of the pion with the perturbative ERBL evolution of the pion DA.
We calculate the meson-photon transition form factors for the $\pi^0$, $\eta$ and $\eta^\prime$ using the hard-scattering formalism.
We point out that a widely-used approximation of replacing $\phi\left (x, (1-x) Q \right)$ with $\phi(x,Q)$ in the calculations
will unjustifiably reduce the predictions for the meson-photon transition form factors.
It is found that the four models of the pion DA discussed give very different predictions for the $Q^2$ dependence of the meson-photon transition form factors
in the region of $Q^2>30$ GeV$^2$.
More accurate measurements of these transition form factors at the large $Q^2$ region will be able to distinguish
different models of the pion DA.
The rapid growth of the large $Q^2$ data for the pion-photon transition form factor reported by the \babar\ Collaboration is difficult to explain within the current framework of QCD.
If the \babar\ data for the meson-photon transition form factor is confirmed, 
it could indicate physics beyond-the-standard model,
such as a weakly-coupled elementary  $C=+$ axial vector or pseudoscalar $z^0$ in the few GeV domain,  an elementary field which would provide the coupling
$\gamma^* \gamma \to z^0 \to \pi^0$ at leading twist.
Our analysis thus indicates the importance of additional measurements of the  pion-photon transition form factor at large $Q^2$.

\end{abstract}


\maketitle

\section{Introduction}

The \babar\ Collaboration has reported measurement of the photon to  pseudoscalar-meson transition form factors from the $\gamma^* \gamma \to M$ process
for the $\pi^0$ \cite{BaBar_pi0}, $\eta_c$ \cite{BaBar_etac}, $\eta$, and $\eta^\prime$ \cite{BaBar_eta,Druzhinin10}.
The momentum transfer $Q^2$ range covered by the \babar\ experiments is much larger than the range studied by the CELLO \cite{CELLO} and CLEO \cite{CLEO} collaborations.
More significantly, the \babar\ data for the $\pi^0$-$\gamma$ transition form factor
exhibit a rapid growth for $Q^2 > 15$ GeV$^2$ which is unexpected from QCD calculations, whereas the data for the other 
transition form factors agree well with previous measurements and theoretical calculations.

QCD computations for exclusive processes are considerably more subtle than inclusive processes since one deals with 
hadron dynamics at the amplitude level.
The foundation for calculating exclusive processes  at high momentum transfer in QCD was laid down almost 30 years
ago \cite{LepageB,EfremovR, DuncanM}.
It was shown in~\cite{LepageB} that  the pion electromagnetic form factor and transition form factor (TFF), the simplest exclusive processes
involving the strong interaction, can be calculated as a convolution of a perturbatively-calculable hard scattering amplitude (HSA),
and the gauge-invariant meson distribution amplitude 
(DA) which incorporates the nonperturbative dynamics of the QCD bound-state.
The distribution amplitude, $\phi(x,Q)$ is the $q  \bar q$ light-front wavefunction (LFWF) $\psi(x, \mbf{k}_\perp)$,
 the eigenstate of the QCD light-front Hamiltonian
 in light-cone gauge, integrated over transverse momenta
$\mbf{k}_\bot^2 \le Q^2$.  Here $x= k^+/P^+ = (k^0+ k^z)/(P^0+ P^z)$ is the light-front momentum fraction of the quark. 
The DA has the physical interpretation as the amplitude to find constituents with longitudinal light-front momentum $x$ and $1-x$ 
in the pion which are non-collinear up to the scale $Q$.
The form of the DA can be confronted with the results of various processes
sensitive to the form of the DA and calculated using non-perturbative methods \cite{BrodskyLPRD1808}.
There are also important constraints from the lowest moments of the pion DA obtained from 
lattice gauge theory~ \cite{QCDSF/UKQCD_Braun06,RBC+UKQCD_Arthur10}.

The evolution of the pion DA is governed by the Efremov-Radyushkin-Brodsky-Lepage (ERBL)
equation~\cite{LepageB,EfremovR,DuncanM}.
The form of the pion DA: $\phi^{asy}(x)  =\sqrt 3  f_\pi x(1-x) $  and the resulting predictions for  elastic and transition form factors at 
the asymptotic limit  $Q^2 \ra \infty$  can be predicted from first principles~\cite{LepageB}.  The results are independent of the input form of 
the distribution amplitude at finite $Q^2.$   
However, the prediction for the elastic form factors using the `asymptotic' form for the pion DA
at finite $Q^2$ range are not  successful when compared with available experimental data. 
This has led to many theoretical investigations of the shape of  the pion DA at low $Q^2$ which reflect the bound state dynamics.
Some models which are vastly different from the asymptotic form have been suggested; however,  these forms 
lack physical motivation and contradict the lattice constraints.
This is manifested by the suggestion of a `flat' form~\cite{ADorokhov09,flat,Polyakov09} for the pion DA in order to explain
the recent \babar\ measurements~\cite{BaBar_pi0} for the pion TFF.

The effects associated with the transverse momentum degree of freedom  have been analyzed in Refs. \cite{CaoHM96,JakobKR96,KrollR96}.
It was shown in \cite{CaoHM96} that the transverse momentum dependence in both the HSA and
the LFWF needs to be considered in order to make predictions for the pion TFF for $Q^2$ of the order of a few GeV$^2$.

The pion form factor has been calculated using the asymptotic DA and Chernyak-Zhitnitsky (CZ) form \cite{CZ}
at next-to-leading order (NLO) \cite{MelicNP02,AguilaC81,Braaten83,KMR86}, using the standard hard-scattering approach when
the $\mbf{k}_\bot$-dependence in the HSA is ignored.
The next-to-next-to-leading order corrections to the hard-scattering amplitude were calculated in~\cite{MelicMP03}
using the conformal operator product expansion.
The form factor has also been studied \cite{JakobKR96,KrollR96} using the modified hard scattering approach in which
the $\mbf{k}_\perp$-dependence is considered together with gluon radiative corrections.
In these calculations the evolution effects were often shown together with
high order corrections. However, due to the limitation on the form used for the pion DA,
the effects from evolution have not been fully explored.
There are many other theoretical studies of the pion-photon transition form factors (see for example
\cite{BMS,BaBar_expln_LiM09,WuH10,Kroll10,BroniowshiA10,GorchetinGS11,PhamP11,ADorokhov10,SAgaevBOP11,ZuoJH10,StofferZ11,KotkoP09,MikhailovS09,RobertsRBGT10,BMPS11,WuH11}).

In this paper, we reexamine the relation between the light-front wavefunction and the distribution amplitude
and calculate the meson-photon transition form factors for the $\pi^0$, $\eta$ and $\eta^\prime$.
Various forms of the meson distribution amplitude and their evolution are studied in Section II. 
Our analysis integrates the nonperturbative bound state dynamics of the pion predicted by light-front holographic QCD with the perturbative QCD ERBL
evolution of the pion distribution amplitude, thus extending the applicability of  AdS/QCD results to large $Q^2$.
The pion-photon transition form factors for the real and virtual photons are calculated in Section III.
The $\eta$-photon and $\eta^\prime$-photon transition form factors are studied in Section~IV.
Some conclusions are given in the last section.

\section{Pion light-front wavefunction and distribution amplitude} 

The pion distribution amplitude in the light-front formalism~\cite{LepageB} is the integral of the 
valence $q \qbar$ light-front wavefunction (LFWF) in light-cone gauge $A^+=0$
\bea
\phi(x,Q)=\int_0^{Q^2}\frac{ d^2 \mbf{k}_\bot}{16 \pi^3} \psi_{q \bar q/ \pi}(x, \mbf{k}_\bot).
\label{eq:DALC}
\eea
The pion DA can also be defined  in terms of the matrix element of the axial isospin current  between a physical pion and the vacuum state~\cite{BrodskyDFL86}
\bea
\phi(x,Q)=\int \frac{d z^-}{2 \pi}{\rm e}^{i (2 x -1) z^-/2}
\left \langle 0 \left| \bar \psi (-z) \frac{\gamma^+ \gamma_5}{2 \sqrt{2}} \Omega \, \psi(z) \right| \pi \right \rangle^{(Q)}_{z^+=z_\perp=0; \, p_\pi^+=0},
 \eea
where 
\bea
\Omega={\rm exp} \left\{ i g \int^{1}_{-1} d s A^+ (z s) z^{-}/2 \right\},
\eea
is a path-ordered factor making $\phi(x,Q)$ gauge invariant.
The pion DA satisfies the normalization condition derived from considering the decay process $\pi \ra \mu \nu$ ($N_C = 3$)
\bea
\int_0^1 {\rm d}x \, 
\phi(x,\mu)  = \frac{f_\pi}{2\sqrt{3}},
\label{Eq:DAnormalization}
\eea
where  $f_\pi = 92.4$ MeV is the pion decay constant and $\mu$ is an arbitrary scale.

By definition the bound state LFWF $\psi_{q \bar q/ \pi}(x, \mbf{k}_\bot)$  
has important support only when the virtual states are near the energy shell, {\it i.e.}
\bea
\varepsilon^2=\left | m_\pi^2-\frac{\mbf{k}_\bot^2+m_q^2}{x(1-x)}\right | < \mu_F^2,
\eea
where $\mu_F$ can be viewed as the factorization scale.
Thus a `cut-off' on the transverse momentum is implied in the definition for the soft component of  the LFWF:  $\psi^{\rm soft}_{q \bar q/ \pi}(x, \mbf{k}_\bot)$.
A natural  way to implement this cut-off is to require the  LFWF to decrease quickly
for large $\mbf{k}^2_\bot$, for example, via an exponential function as first suggested in the model discussed in \cite{BrodskyHL}.
One can write a parameterization form for the LFWF as in \cite{JakobK93} 
\bea
\psi_{q \bar q/ \pi}^{\rm soft}(x, \mbf{k}_\bot) &\equiv& \phi(x) \, \Sigma (x,\mbf{k}_\bot) \nonumber \\
&=& \phi(x) \frac{8 \pi^2}{\kappa^2}\frac{1}{x(1-x)} {\rm exp}\left(-\frac{\mbf{k}_\bot^2}{2 \kappa^2 x (1-x)} \right),
\label{eq:SLFWF}
\eea
where $\kappa$ is the  gap parameter, and
\bea 
\phi(x)=\int_0^\infty\frac{ d^2 \mbf{k}_\bot}{16 \pi^3} \psi^{\rm soft}_{q \bar q/ \pi}(x, \mbf{k}_\bot),
\label{eq:phis}
\eea
and the function $\Sigma$ satisfies\footnote{Strictly speaking a cut-off of $ |\mbf{k}^2_\bot|^{\rm max} \sim x(1-x) 
\mu^2_F$
is still in place for the soft wave function given by Eq.~(\ref{eq:SLFWF}). However, calculations are not sensitive 
to this cut-off due to the nature of rapid decreasing of the wave function. 
Thus it is commonly expressed in the literature that $|\mbf{k}^2_\bot|^{\rm max}= \mu^2_F$, or $ |\mbf{k}^2_\bot|^{\rm max}= \infty$.},
\bea
\int_0^\infty\frac{ d^2 \mbf{k}_\bot}{16 \pi^3} \Sigma (x,\mbf{k}_\bot)=1.
\eea

A common practice used in the literature in determining the parameter $\kappa$ is calculating the non-perturbative properties of the pion 
and comparing with the experimental measurements of  these quantities.
However, this process only allows one to constrain 
 $\kappa$ in a relative large range due to the uncertainty 
of  the experimental measurements.
For example, the root of the mean square transverse momentum of the valence quarks, defined as 
\bea
\sqrt{\left< \mbf{k}_\bot^2 \right>} &=&\left( \frac{1}{P_{q {\bar q}}}  \int_0^1 d x \int_0^\infty\frac{ d^2 \mbf{k}_\bot}{16 \pi^3} \mbf{k}_\bot^2 
\left| \psi^{\rm soft}_{q \bar q/ \pi}(x, \mbf{k}_\bot) \right|^2 \right)^{1/2},
\label{eq:mskT}
\eea
where $P_{q {\bar q}}$ is the probability of the valence Fock state of the pion
\bea
P_{q {\bar q}}=\int_0^1 d x \int_0^\infty\frac{ d^2 \mbf{k}_\bot}{16 \pi^3} \left| \psi^{\rm soft}_{q \bar q/ \pi}(x, \mbf{k}_\bot) \right|^2,
\label{eq:Pqqbar}
\eea
is estimated to be in the range of $300 \sim500$ MeV from experimental measurement on the charge radius of the pion. 
Thus $\kappa$ is not well determined by Eq.~(\ref{eq:mskT}).

Brodsky, Huang, and Lepage \cite{BrodskyHL} obtained a constraint for the soft LFWF at $\mbf{k}_\bot=0$, $\psi_{q \bar q/ \pi}^{\rm soft}(x, \mbf{k}_\bot=0)$,
by studying the decay of $\pi^0 \ra \gamma \gamma$. However, we note that the decay  $\pi^0 \ra \gamma \gamma$ is a long-distance process for which the higher Fock states
should make substantial contributions as well, 
since there are extra interactions with the quark propagator between the two photons which vanish at high $Q^2$ in the light-cone gauge. 
Therefore $\kappa$ cannot be well determined from the constraints imposed by the decay process.

From these considerations we will treat $\kappa$ in Eq.~(\ref{eq:SLFWF}) as a phenomenological parameter which is allowed to change in a certain range.
It is equivalent to treat the probability of the valence Fock state, $P_{q \qbar}$, or the root of the mean square transverse momentum of the valence quarks,
$\sqrt{\left< \mbf{k}_\bot^2 \right>}$, as a parameter. 

The LFWF $\psi_{q \bar q/ \pi}(x, \mbf{k}_\bot)$ in Eq.~(\ref{eq:DALC})
contains all the non-perturbative information of the pion.
There are also perturbative corrections that behave as $\alpha_s(\mbf{k}^2_\perp)/\mbf{k}^2_\perp$ for large 
$\mbf{k}^2_\perp$,
coming from the  fall-off of the LFWF $\psi(x, \mbf{k}_\perp)$ due to hard gluon radiation  \cite{LepageB,BrodskyHL}.
Both soft and hard regimes are important to compute the pion transition form factor for all values of $Q^2$.

\subsection{Soft Evolution of the Pion Distribution Amplitude}

Substituting Eq.~(\ref{eq:SLFWF}) into Eq.~(\ref{eq:DALC}) one obtains
\bea
\phi(x,Q) = \phi(x) \left[ 1- {\rm exp}\left(-\frac{Q^2}{2 \kappa^2 x (1-x)} \right) \right] ,
\label{eq:DA1}
\eea
where $\phi(x)$ is given by Eq.~(\ref{eq:phis}).
Eq.~(\ref{eq:DA1}) gives a factorization model for the $Q^2$ dependence of the distribution amplitude in the soft domain.
The soft $Q^2$ dependence in Eq.~(\ref{eq:DA1}) can be safely ignored for $Q  > 1$ GeV
for the typical values of $\kappa \sim 0.5-1.0$ GeV.  In the regime of $Q > 1$ GeV one needs to consider the hard gluon exchanges that 
provide additional logarithmic $Q^2$ dependence in $\phi(x,Q)$, as given by the ERBL evolution equation discussed below.

Many efforts have been made in determining the pion DA at a low momentum transfer scale $\mu_0 \sim 0.5 - 1$ GeV.
Most of these studies concentrate on the determination of the first few terms in the solution of the evolution equation
for the pion DA discussed in the next section.
However the pion DA at a low scale could differ significantly from its asymptotic form due to the slow convergence
of the evolved DA. Using only a few terms of the full solution will put a strong limitation on the studies.
The following forms have been suggested.

\begin{enumerate}[(a)]

\item
The asymptotic form~\cite{LepageB,BrodskyDFL86,Rady04}
\bea 
\phi^{\rm asy}(x)=\sqrt{3} f_\pi x (1-x).
\eea

\item
The AdS/QCD form \cite{Brodsky:2006uqa, Brodsky:2007hb}
\bea
\phi^{\rm AdS}(x)=\frac{4}{\sqrt{3} \pi} f_\pi \sqrt{x (1-x)}.
\eea

\item
The Chernyak-Zhitnitsky  \cite{CZ} form
\bea
\phi^{\rm CZ}(x)&=&5 \sqrt{3} f_\pi x (1-x) (1- 2 x)^2 \nonumber \\
&=& \sqrt{3} f_\pi x (1-x) \left[ 1+\frac{2}{3} C_2^{(2/3)}(1- 2 x) \right].
\eea

\item
The `flat' form \cite{Polyakov09}
\bea
\phi^{\rm flat}(x)=\frac{f_\pi}{2 \sqrt{3}} \left[ N+6 (1-N) x (1-x) \right].
\eea
\end{enumerate}

The DA model (b) follows from the precise mapping of string amplitudes in Anti-de Sitter (AdS) space to the light-front wavefunctions
of hadrons in physical space-time using holographic methods~\cite{Brodsky:2006uqa, Brodsky:2007hb, Brodsky:2008pf, deTeramond:2008ht}.
However, an extended AdS model
 with a Chern Simons action maps to the asymptotic DA form  $x(1-x)$ \cite{Grigoryan:2008up} rather than the AdS form $\sqrt{x(1-x)}$.
A discussion of the pion form factor is discussed in the framework of light-front holographic mapping in a forthcoming paper \cite{BrodskyCTAdS11}.
Model (c) was suggested on the basis of a calculation using  QCD sum rules
and model~(d) was advocated in explaining the recent \babar\ data for the pion TFF \cite{BaBar_pi0}. 
The end-point non-vanishing models, similar to model (d), were also obtained \cite{ArriolaB02-06,BzdakP03,DorokhovB06} 
for the pion and photon DAs using chiral quark models and  Regge models before the \babar\ results were reported.
Normally one expects that the light-front wavefunction of a composite hadron to vanish at the $x=0,1$ end-points to ensure a finite expectation value of the kinetic energy operator.
A set of  pion DAs (termed the BMS models) including only the first two terms in the general solution of the ERBL 
(see Eq.~\ref{eq:DAQ2}) below)
were proposed \cite{BMS} by comparing the light-cone sum rules calculations for the pion TFF with the CELLO and CLEO data. 
Theoretical calculations using transverse lattice gauge theory with discrete light cone quantization  \cite{DalleyS03} and chiral quark models \cite{BroniowskiAG08}
generally suggest that that the pion DA is considerably broader than the asymptotic form.

\begin{figure}[h]		
\includegraphics[width=80mm]{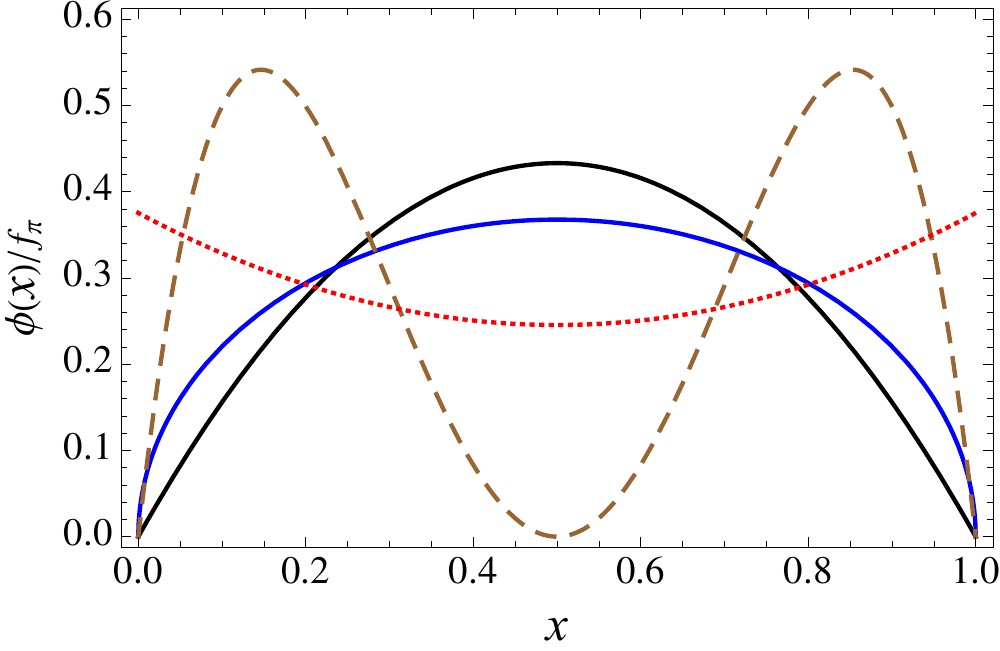}
\caption{
The four commonly used models for the pion distribution amplitude. The curves from bottom to top at $x=0.5$ are for the CZ, `flat', AdS/QCD, and asymptotic forms, respectively. \label{DAmodels}}
\end{figure}

The four models 
are shown in Fig.~\ref{DAmodels}.
Model~(d) is not actually only flat over the whole range of $x$ -- it is end-point enhanced.
Models (c) and (d) have very different shape from (a) and (b).
The zero-value of the CZ form in the middle point ($x=0.5$), where the pion momentum is shared equally between the quark and the antiquark,
and the enhancement of the `flat' form in the end-points ($x=0,1$), where the pion momentum is mostly carried by the quark or the antiquark,
are hard to understand in terms of the bound state dynamics of the pion. The zero-value of the CZ form in the middle point also disagrees with
the estimation using the QCD sum rule method reported in \cite{BraunF89}, 
$\phi(x=0.5)=(0.17 \pm 0.03) f_\pi$.

Using the models of the pion DA discussed above we can construct the corresponding LFWF from Eq.~(\ref{eq:SLFWF}).
The LFWF constructed with the `flat' form (model (c)) for the pion DA is non-normalizable since
the probability of finding the valence Fock state in the pion (Eq.~(\ref{eq:Pqqbar})) becomes infinity \cite{SAgaevBOP11}.
We list the values for the gap parameter
 $\kappa$ and the root of the mean square transverse momentum of the valence quarks
$\sqrt{\left< \mbf{k}_\bot^2 \right>}$ for the three choices of the probability $P_{q \bar q}=0.25,\, 0.50$
and $0.80$ in Table \ref{Table:kappa}.
For the `flat' model Eq.~(\ref{eq:Pqqbar}) is divergent so we adopt $\kappa^2=0.530$ GeV$^2$ \cite{flat} and $N=1.3$ \cite{Polyakov09}, which were chosen to explain the \babar\ data.

\begin{table}
\caption{Properties of the soft light-front wavefunction corresponding to various models of the pion DA. \label{Table:kappa}}
\begin{tabular}{|c|c|c|c|}
\hline
 $P_{q \bar q}$ & DA & $\kappa$ (GeV) & $\sqrt{\left< \mbf{k}_\bot^2 \right> }$ (GeV)\\ \cline{1-4}
 {\multirow{3}{*}{0.25}} & $\phi^{\rm asy} (x)$ & 0.826 & 0.370 \\ \cline{2-4}
  & $\phi^{\rm AdS}(x)$ & 0.859 & 0.350 \\ \cline{2-4}
 & $\phi^{\rm CZ}(x)$ & 1.210 & 0.403 \\ \hline
 {\multirow{3}{*}{0.50}} & $\phi^{\rm asy} (x)$ & 0.584 &0.261 \\ \cline{2-4}
  & $\phi^{\rm AdS}(x)$ & 0.607 & 0.248 \\ \cline{2-4}
 & $\phi^{\rm CZ}(x)$ & 0.855 & 0.285 \\ \hline
 {\multirow{3}{*}{0.80}} & $\phi^{\rm asy} (x)$ & 0.462 &0.207 \\ \cline{2-4}
  & $\phi^{\rm AdS}(x)$ & 0.480 & 0.196 \\ \cline{2-4}
 & $\phi^{\rm CZ}(x)$ & 0.676 & 0.225 \\ \hline

 \end{tabular}
 \end{table}

\subsection{Hard Evolution of the Pion Distribution Amplitude}

The evolution of the pion DA at large $Q$ is governed by the ERBL equation.
The solution to the ERBL equation can be expressed~\cite{LepageB,EfremovR}
in terms of Gegenbauer polynomials,
\bea
\phi(x,Q)= x (1-x)  \sum_{n=0,2,4, \cdots}^{\infty} a_n(Q)  C_n^{3/2}(2 x -1),
\label{eq:DAQ2}
\eea
where 
\bea
a_n(Q) = \left[\frac{\alpha_s(\mu_0^2)}{\alpha_s(Q^2)} \right]^{\gamma_n/\beta_0} a_n(\mu_0) ,
\label{eq:aQ}
\eea
 at leading order. The coefficients
 $\{a_n(\mu_0)\}$ are the coefficients in the Gegenbauer expansion of the DA at the initial scale $\mu_0$, 
\bea
\phi(x,\mu_0)= x (1-x) \sum_{n=0,2,4, \cdots}^{\infty} a_n(\mu_0) C_n^{3/2}(2 x -1),
\label{eq:DAmu0}
\eea
and follow from the orthonormality of the Gegenbauer polynomials
\bea
a_n(\mu_0)=\frac{4(2 n+3)}
{(n+2)(n+1)} \int_{0}^1{\rm d} x \phi(x,\mu_0) C_n^{3/2}(1-2 x).
\label{eq:anmu0}
\eea 

The QCD coupling constant $\alpha_s(Q^2)$ is taken to have the leading-order form
\bea 
\alpha_s(Q^2)=\frac{4 \pi}{\beta_0 {\rm ln}\left( Q^2/\Lambda_{\rm QCD}^2 \right)},
\eea 
where $\Lambda_{\rm QCD}$ is the QCD scale parameter and $\beta_0$ is the QCD beta function one-loop coefficient
$\beta_0=11- \frac{2}{3} \, n_f$.
The anomalous dimensions $\gamma_n$ appearing in Eq. (\ref{eq:aQ}) 
\bea
\gamma_n=\frac{4}{3} \left[3+\frac{2}{(n+1)(n+2)}-4 \sum_{j=1}^{n+1}\frac{1}{j} \right] ,
\eea
are the eigenvalues of the evolution kernel~\cite{LepageB,EfremovR}.
The coefficient $a_0(\mu_0)=\sqrt{3} f_\pi$ for any model of the pion DA, since the pion DA should satisfy the normalization condition
Eq.~(\ref{Eq:DAnormalization})
with $C_0^{3/2}(z)=1$, and $\int_0^1 dx x(1-x) C_n^{3/2}(1-2 x)=0$ for $n \geq 2$.

The coefficients $a_n(\mu_0)$ are computed at the initial  scale $\mu_0 = 1$ GeV (where the effects of hard gluons is negligible and the scale dependence of the
soft evolution is not important). Thus we choose the initial condition $\phi(x, \mu_0 \simeq 1~ {\rm GeV} ) \simeq \phi(x)$, with $\phi(x)$  given by Eq. (\ref{eq:phis}).
At leading order the asymptotic form does not evolve since all the expansion coefficients $\{a_n(\mu_0) \}$, but $a_0(\mu_0)=\sqrt{3} f_\pi$,
vanish for model (a). The coefficients $\{ a_n(\mu_0) \}$ for model (c) (the CZ form) are very simple since the model
essentially includes only the first two terms in the Gegenbauer polynomials, $a_0(\mu_0) =\sqrt{3} f_\pi$ and $a_2(\mu_0) =2/\sqrt{3} \, f_\pi$.
For model (b) (the AdS form) we include the
first 50 terms ({\it i.e.} up to $n=100$) in Eqs.~(\ref{eq:DAQ2}) and (\ref{eq:DAmu0}) in our calculation.
The first 10 values of $a_n({\mu_0})$ for the AdS model and the nonzero coefficients for the asymptotic and CZ models 
are listed in Table \ref{Table_anmu0}.
It was found that for the AdS model the calculation with 51 terms only brings 
a few percent corrections to the calculation with 21 terms over a large range of $x$.

\begin{table}
\caption{The coefficients $a_n(\mu_0)$ for the asymptotic, AdS and CZ models for the pion DA. \label{Table_anmu0}}
\begin{tabular}{|c|c|c|c|c|c|c|c|c|c|c|c|}
\hline
 \multicolumn{2}{|c|}{n} & 0 & 2& 4& 6 & 10 & 12 & 14 & 16 & 18 & 20  \\ \cline{1-12}
{\multirow{3}{*}{$a_n(\mu_0)/a_0(\mu_0)$ }} & AdS& $1$ & 0.1461 & 0.0573 &  0.0305 & 0.0189 & 0.0129 & 0.0094 & 0.0071 & 0.0056 & 0.0045  \\  
\cline{2-12}
& CZ & $~1~$ &  2/3 & \multicolumn{8}{c|}{0 for n $\geq$ 4} \\ \cline{2-12}
 & asy & $1$ & \multicolumn{9}{c|}{0 for n $\geq$ 2} \\ \hline
 \end{tabular}
 \end{table}
 
 It is problematic to expand the `flat' DA in term of the Gegenbauer polynomials at the initial scale $\mu_0$, since the expansion Eq.~(\ref{eq:DAmu0})
 converges if, and only if, $\phi(x,\mu_0)$  vanishes at end-points \cite{LepageB,HTF}.
We will not try to apply the ERBL equation to the `flat' DA, but just make the note that if one applied the ERBL equation to the `flat' DA,
one would enforce the suppression at the end-points as soon as the evolution starts.

\begin{figure}[h]	
\includegraphics[width=80mm]{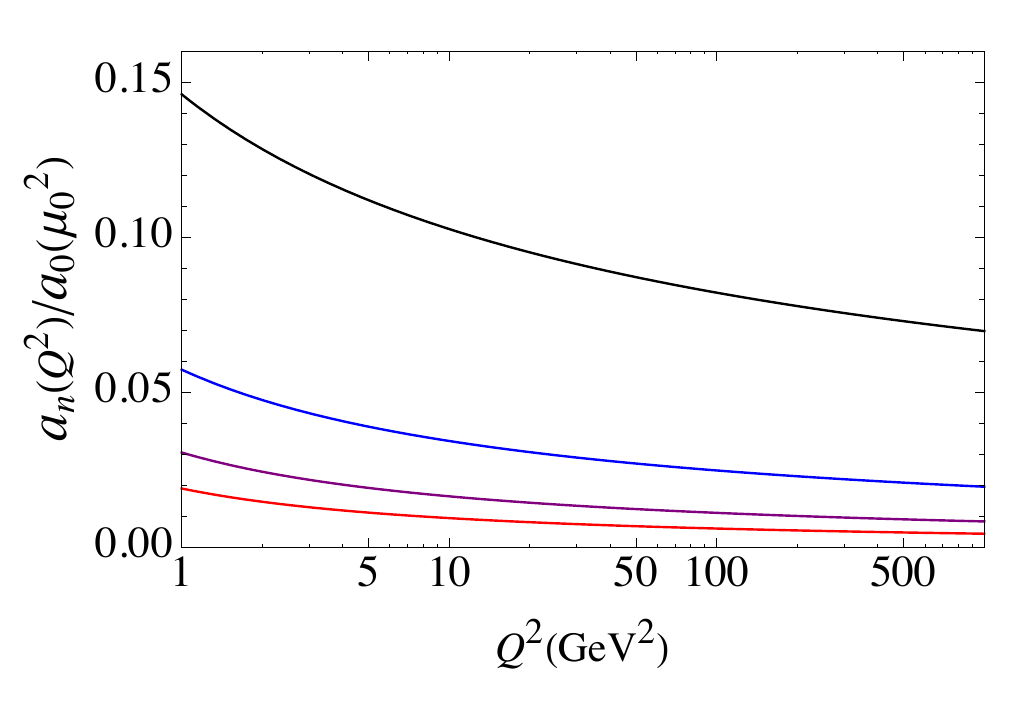}
\caption{
 Evolution of the expansion coefficients $a_n(Q^2)$ for the AdS distribution amplitude. The curves from top to bottom are for
 $n=2,4,6$ and $8$ respectively.
 \label{anQ2}}
\end{figure}

\begin{figure}[h]		
\includegraphics[width=80mm]{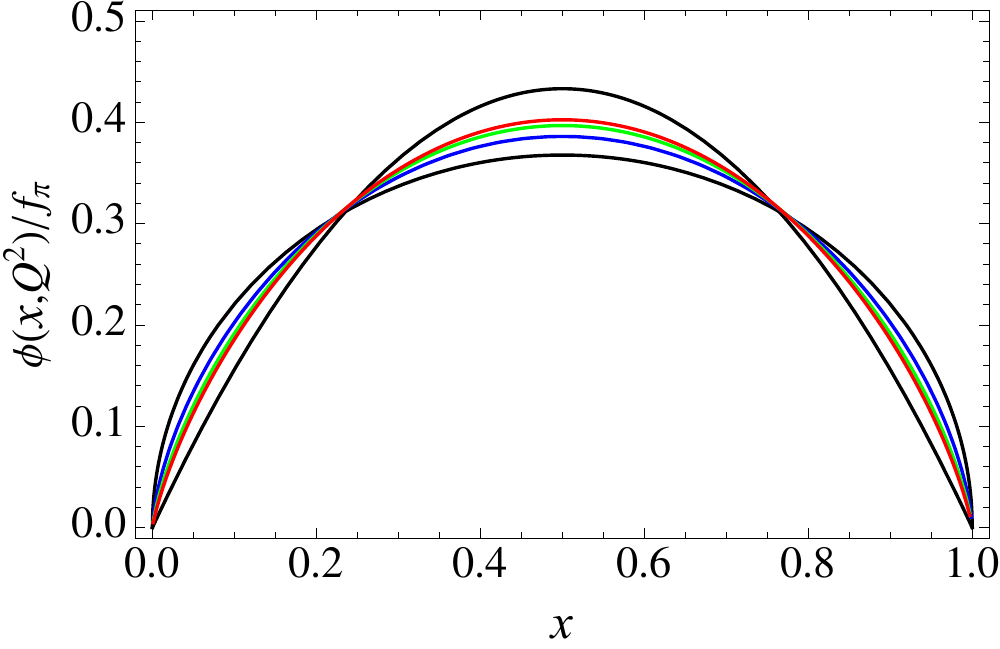}
\caption{
Evolution effects shown for the AdS model for the pion DA.
The curves from bottom to top at $x=0.5$ are for $Q^2= 1,\,10,\, 100$ and $1000$ GeV$^2$,
and the asymptotic DA, respectively.}
 \label{DA_AdS_evolved}
\end{figure}

\begin{figure}		
\begin{center}
\includegraphics[width=80mm]{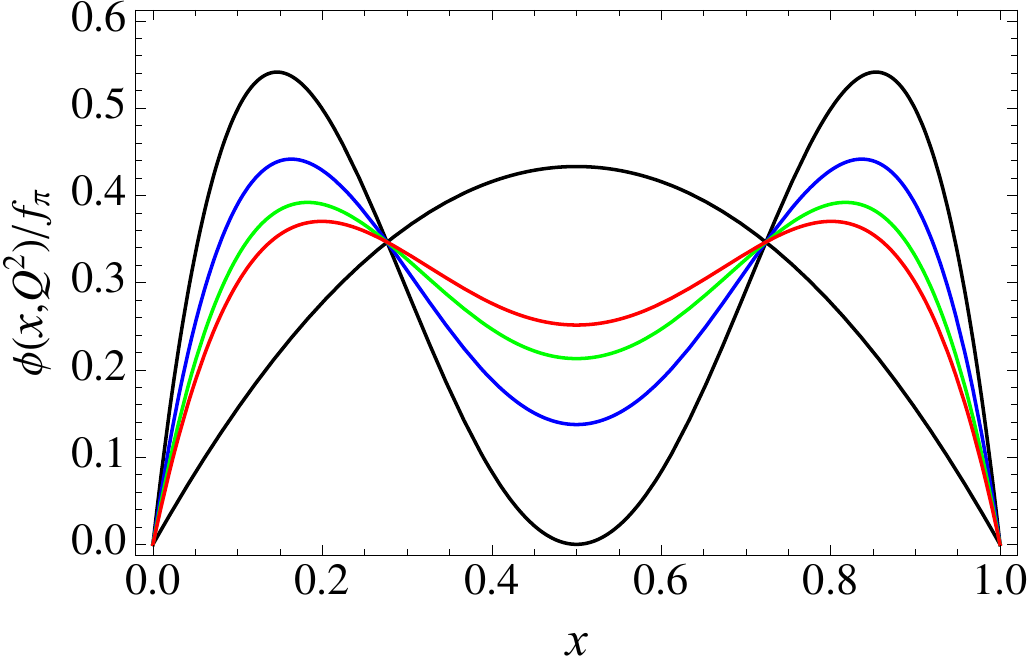}
\caption{
 Similar as in Fig. \ref{DA_AdS_evolved} but for the CZ model for the pion DA.}
\label{DA_CZ_evolved}
\end{center}
\end{figure}

 The first term in Eq.~(\ref{eq:DAQ2}) represents the asymptotic form of the pion DA and the asymptotic form does not  evolve with  $Q^2$.
 The other distribution amplitudes have Gegenbauer polynomial components with nonzero anomalous
dimensions which drive their contributions to zero for large values of $Q$.  One can start with any distribution amplitude $\phi(x, \mu_0)$
at any finite scale and expand it as $x(1-x)$ times Gegenbauer polynomials. Only its projection on the lowest Gegenbauer polynomial with 
zero anomalous moment survives. This is illustrated in Fig.~\ref{anQ2} for the first few expansion coefficients $a_n(Q^2)$ for the AdS distribution amplitude.
The evolution effects of the DA at leading order are shown in Figs.~\ref{DA_AdS_evolved} and \ref{DA_CZ_evolved}
for the AdS model and CZ model for the pion DA respectively.
In our numerical calculations we used $\mu_0=1$ GeV and $\Lambda_{\rm QCD}=225$ MeV.
Performing evolution at NLO modifies the results slightly.
It can be seen that evolution effects change the shape of the CZ form significantly,
while the effect on the AdS form is not as dramatic. 
In the asymptotic $Q^2 \to \infty$ limit  the asymptotic DA is recovered. 

\subsection{Moments of the Pion Distribution Amplitude}

Important constraints for the form of the distribution amplitudes also follow from QCD lattice computations.
The latest results for the second moment of the pion DA,
\bea
\langle \xi^2\rangle_{\mu^2} = \frac{\int^1_{-1} d \xi  \xi^2 \phi(\xi,\mu^2)}{\int^1_{-1} d \xi \phi(\xi ,\mu^2)} ,
\eea
where $\xi=1-2 x$, are 
$\langle \xi^2\rangle_{\mu^2 =  4\, {\rm GeV}^2} = 0.269 \pm 0.039$ \cite{QCDSF/UKQCD_Braun06}
and $0.28 \pm 0.03$ \cite{RBC+UKQCD_Arthur10}.
The second moments calculated at the initial scale $\mu_0$ for the  four models of the pion DA described above are 
$0.20$ (asymptotic), $0.25$ (AdS/QCD), $0.43$ (CZ) and $0.37$ (`flat'), respectively.
Using the ERBL evolution equations we can compute the second moments at the scale $\mu^2 = 4$ GeV$^2$.
We find the values $0.20$ (asymptotic), $0.24$ (AdS/QCD) and $0.38$ (CZ).
The agreement between the AdS model value and the lattice results is better than the result found for the asymptotic model and CZ model.
We also note that the measurement of the pion DA in diffractive di-jet production reported by the E791 Collaboration \cite{E791_Aitala01}
supports a centrally-peaked DA such as  the asymptotic and AdS/QCD models.
The second moment alone, while providing important information for the pion DA, will not put strong constraints on the shape of the pion DA,
since it is a quantity obtained by integrating the DA over the whole range of $x$.

\section{Pion-photon transition form factors}

The pion-photon transition form factor can be extracted from the two-photon process $\gamma^* (q_1) \gamma^* (q_2) \ra \pi^0$.
When both photons are off-shell with virtuality $Q_1^2=-q_1^2$ and $Q_2^2=-q_2^2$,
the form factor is denoted as $F_{\pi\gamma^*} (Q_1^2,Q_2^2)$.
In the case of one photon being on mass-shell the form factor is denoted as $F_{\pi\gamma} (Q^2)$.

\subsection{Leading order results}

Brodsky and Lepage~\cite{LepageB,BrodskyLPRD1808} predicted the behavior of $F_{\pi \gamma}(Q^2)$
at leading order of $\alpha_s(Q^2)$ and leading twist as
\bea
Q^2 F_{\pi \gamma}(Q^2)=\frac{4}{\sqrt{3}} \int_0^1  {\rm d} x \frac{\phi(x,{\bar x} Q)}{\bar x}
\left[ 1+ O \left(\alpha_s,\frac{m^2}{Q^2} \right) \right],
\label{eq:TFLB1}
\eea
where $x$ is the longitudinal momentum fraction of the quark struck by the virtual photon in the hard scattering process
and ${\bar x}=1-x$ is the longitudinal momentum fraction of the spectator quark.
It was argued~\cite{LepageB} that the boundary condition of DAs vanishing at the end-points
faster than $x^\epsilon$ for some $\epsilon > 0$ would enable one to replace $\phi(x,{\bar x} Q)$ by $\phi(x,Q)$ 
since the difference is non-leading\footnote{It was actually pointed out that the replacement ${\bar x} Q \ra Q/2$
is more appropriate.}, 
and Eq.~(\ref{eq:TFLB1}) becomes
\bea
Q^2 F_{\pi \gamma}(Q^2)=\frac{4}{\sqrt{3}} \int_0^1  {\rm d} x \frac{\phi(x,Q)}{\bar x}
\left[ 1+ O \left(\alpha_s,\frac{m^2}{Q^2} \right) \right].
\label{eq:TFLB2}
\eea
The replacement is sound when one is interested in the leading order behavior of the TFF
and particularly for the behavior at the asymptotic limit $Q^2 \ra \infty$, which is one of the main
purposes  of Ref.~\cite{LepageB}.
However, this approximation is not justified for the calculation
at finite $Q^2$ region where one needs to take into account the evolution effects and NLO corrections.
The dominant contributions to the integrals in Eqs.~(\ref{eq:TFLB1}) and (\ref{eq:TFLB2})
come from small $x$ region, {\it e.g.}, 3/4 of the contributions coming from $x\le 0.5$ for the asymptotic DA.
At the same time the evolution changes the shape of the DA more significantly in the small $x$ region.
Thus the calculations with Eq.~(\ref{eq:TFLB1}) involve a much less evolved DA than Eq.~(\ref{eq:TFLB2}). 
The difference between the calculations using Eqs.~(\ref{eq:TFLB1}) and (\ref{eq:TFLB2}) could be sizable.
Unfortunately, Eq.~(\ref{eq:TFLB2}) has been widely used in the literature as the starting point  to calculate high-order corrections to the pion TFF,
see {\it e.g.}, \cite{MelicNP02,AguilaC81,Braaten83,KMR86,MelicMP03}. 
Similar replacement has been done in the study for other exclusive processes.

It is essential to consider the transverse momentum dependence in both the hard-scattering amplitude and the LFWF in order to 
describe the data at finite $Q^2$ \cite{CaoHM96}. Taking into account the $k_\bot$-dependence,
the pion-photon transition form factor is given by \cite{LepageB,CaoHM96}
\bea
F_{\pi \gamma}(Q^2)=\frac{2}{\sqrt{3}} \int_0^1  {\rm d} x
\int_0^\infty \frac{d^2 \mbf{k}_\bot}{16 \pi^3} T_H(x,Q^2,\mbf{k}_\bot) 
\psi_{q \bar q / \pi}(x,\mbf{k}_\bot),
\label{eq:TFCao1}
\eea
where 
\bea
T_H(x,Q^2,\mbf{k}_\bot)=
\frac{\mbf{q}_\bot \cdot (\bar x \mbf{q}_\bot + \mbf{k}_\bot)}
{\mbf{q}_\bot^2 (\bar x \mbf{q}_\bot + \mbf{k}_\bot)^2} +\left[ x \leftrightarrow {\bar x} \right] ,
\label{eq:THCao}
\eea
is the hard scattering amplitude and $\mbf{q}_\bot^2=Q^2$. Using Eq.~(\ref{eq:TFCao1}) and a Gaussian type LFWF 
one can reproduce the curve displayed by the experimental data at low $Q^2$ \cite{CaoHM96}.
With Eqs.~(\ref{eq:TFLB1}) and (\ref{eq:TFLB2}) the calculations will be near constant for all $Q^2$.

Musatov and  Radyushkin have shown~\cite{Rady2} that if  LFWF depends on the transverse momentum only through 
$\mbf{k}_\bot^2$,
Eq.~(\ref{eq:TFCao1}) can be simplified as
\begin{eqnarray}
Q^2F_{\pi \gamma}(Q^2)=\frac{4}{\sqrt{3}} \int_0^1 \frac{ {\rm d} x}{\bar x}
\int_0^{\bar x Q}  \frac{{\rm d}^2 \mbf{k}_\bot}{16 \pi^3} 
\psi_{q \bar q/ \pi}(x, \mbf{k}_\bot^2).
\label{eq:TFRady}
\end{eqnarray}
For the model wavefunction Eq.~(\ref{eq:SLFWF}) we can factor out the $Q^2$ dependence of the DA at low $Q^2$, Eq.~(\ref{eq:DA1}),
and include
the QCD evolution for higher momenta through the ERBL solution of the DA. One obtains\footnote{Enforcing the cut-off 
$\mbf{k}_\bot^2 \le x(1-x)Q^2$ for the soft LFWF discussed in Section~I,
 Eq.~(\ref{eq:TFCao2}) becomes
$Q^2 F_{\pi \gamma}(Q^2)=\frac{4}{\sqrt{3}} \int_0^1 {\rm d}x \frac{\phi(x,{\tilde Q})}{\bar x}
\left[ 1-{\rm exp} \left( - \frac{ {\tilde Q}^2}{2 \kappa^2 x {\bar x}}  \right) \right ]$
where ${\tilde Q}^2 =x  {\tilde x} Q^2$ with ${\tilde x}=  \, {\rm min}\left(x, {\bar x}\right)$.
The two expressions coincide for $x\le 0.5$ and the differences are negligible unless $Q^2<1$ GeV$^2$.
However, for other exclusive processes that are sensitive to the large-$x$ region
the difference may be sizable.}~\cite{Rady2}
\bea
Q^2 F_{\pi \gamma}(Q^2)=\frac{4}{\sqrt{3}} \int_0^1 {\rm d}x \frac{\phi(x,{\bar x} Q)}{\bar x}
\left[ 1-{\rm exp} \left( - \frac{ \bar x Q^2}{2 \kappa^2 x}  \right) \right ].
\label{eq:TFCao2}
\eea
The pion TFF depends on $Q^2$ through the exponential factor 
 and the pion DA.  Since we have explicitly factored out the low $Q^2$-dependence, the distribution amplitude in Eq. (\ref{eq:TFCao2})
 contains only the hard ERBL evolution.
The exponential factor is important, especially for small $\bar x$ and small $Q^2$, thus it controls
the curvature  of $Q^2F_{\pi \gamma}(Q^2)$ vs. $Q^2$ at low $Q^2$. 
The behavior of the pion TFF at high $Q^2$ is determined dominantly by the pion `hard' DA which
should evolve in a logarithmic manner. 
The exponential factor also plays a role to regularize the calculation with the `flat' DA, which otherwise
involves a divergent integral.

Inserting Eq.~(\ref{eq:DAQ2}) into Eq.~(\ref{eq:TFCao2}) we can write the transition form factor as
\bea
Q^2 F_{\pi \gamma}(Q^2)&=&\frac{4}{\sqrt{3}} f_\pi \sum_{n=0,2,4\cdots}^\infty a_n(\mu_0)
\int_{0}^1{\rm d} x \, x C_n^{3/2}(2 x-1) \nonumber \\
& & \left[\frac{\alpha_s(\mu_0^2)}{\alpha_s({\bar x}^2 Q^2)} \right]^{\gamma_n/\beta_0}
\left[ 1-{\rm exp} \left( - \frac{\bar x Q^2}{2 \kappa^2 x}  \right) \right ].
\label{eq:TFCao4}
 \eea
 which displays the soft and hard dependence.
We need to set $\bar xQ=\mu_0$ for $ \bar x Q<\mu_0$, which assures the convergence of Eq.~(\ref{eq:TFCao4}).
Equations (\ref{eq:TFCao2}) and (\ref{eq:TFCao4})
clearly show that the pion TFF at any given $Q^2$ is 
determined  by $\phi(x,\mu_0)$ and
all evolved DAs from $\mu_0$ to $Q$, with the less-evolved DAs providing major contributions.
For example, half of the contributions at $Q \simeq 3~\mu_0 $ come from $\phi(x, \mu_0)$ for
the asymptotic DA, and this ratio is much higher for broad models for the pion DA.
The contributions from $\phi(x,\mu_0)$ remain significant even when $Q \sim 5 \, \mu_0$.
Thus the evolution effect hardly  shows up until $Q^2$ is very large.
On the other hand, if one uses the distribution $\phi(x,Q)$ in Eqs.~(\ref{eq:TFCao2}) and (\ref{eq:TFCao4})
the evolution effect will be overestimated.

\begin{figure}
\includegraphics[width=100mm]{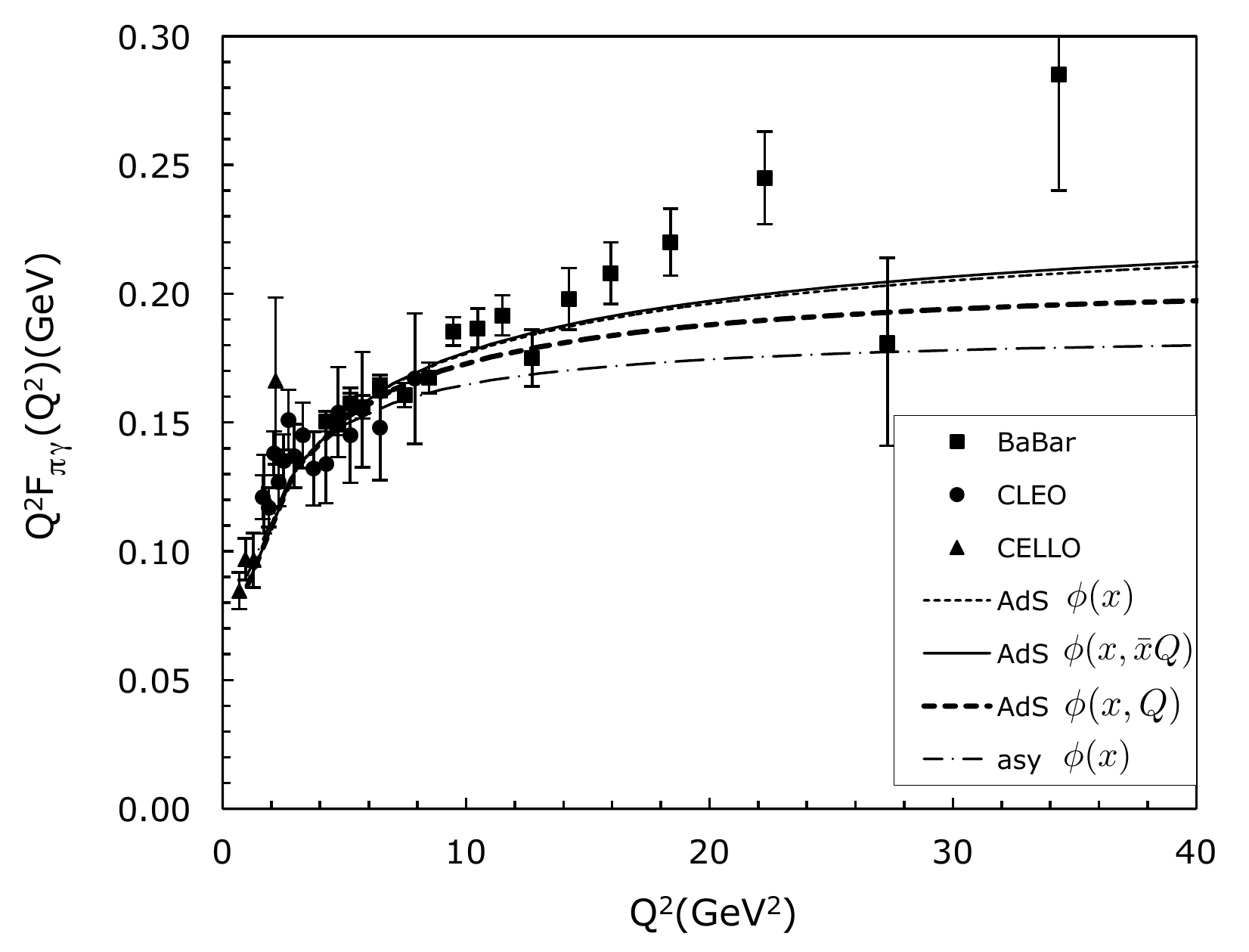}
\caption{
The pion-photon transition form factor shown as $Q^2 F_{\pi \gamma}(Q^2)$
calculated using Eq.~(\ref{eq:TFCao2}) with different prescriptions for $\phi(x,\mu)$:
solid curve  -- $\phi^{\rm AdS}(x,\bar x Q)$,
dashed curve -- $\phi^{\rm AdS}(x)$,
thick-dashed curves -- $\phi^{\rm AdS}(x,Q)$,
and dash-dotted curve -- $\phi^{\rm asy}(x)$. $P_{q \bar q}=0.50$.
The data are taken from \cite{BaBar_pi0,CELLO,CLEO}.\label{fig:Q2FpiTF_LO_AdS}}
\end{figure}

\begin{figure}		
\begin{center}
\includegraphics[width=100mm]{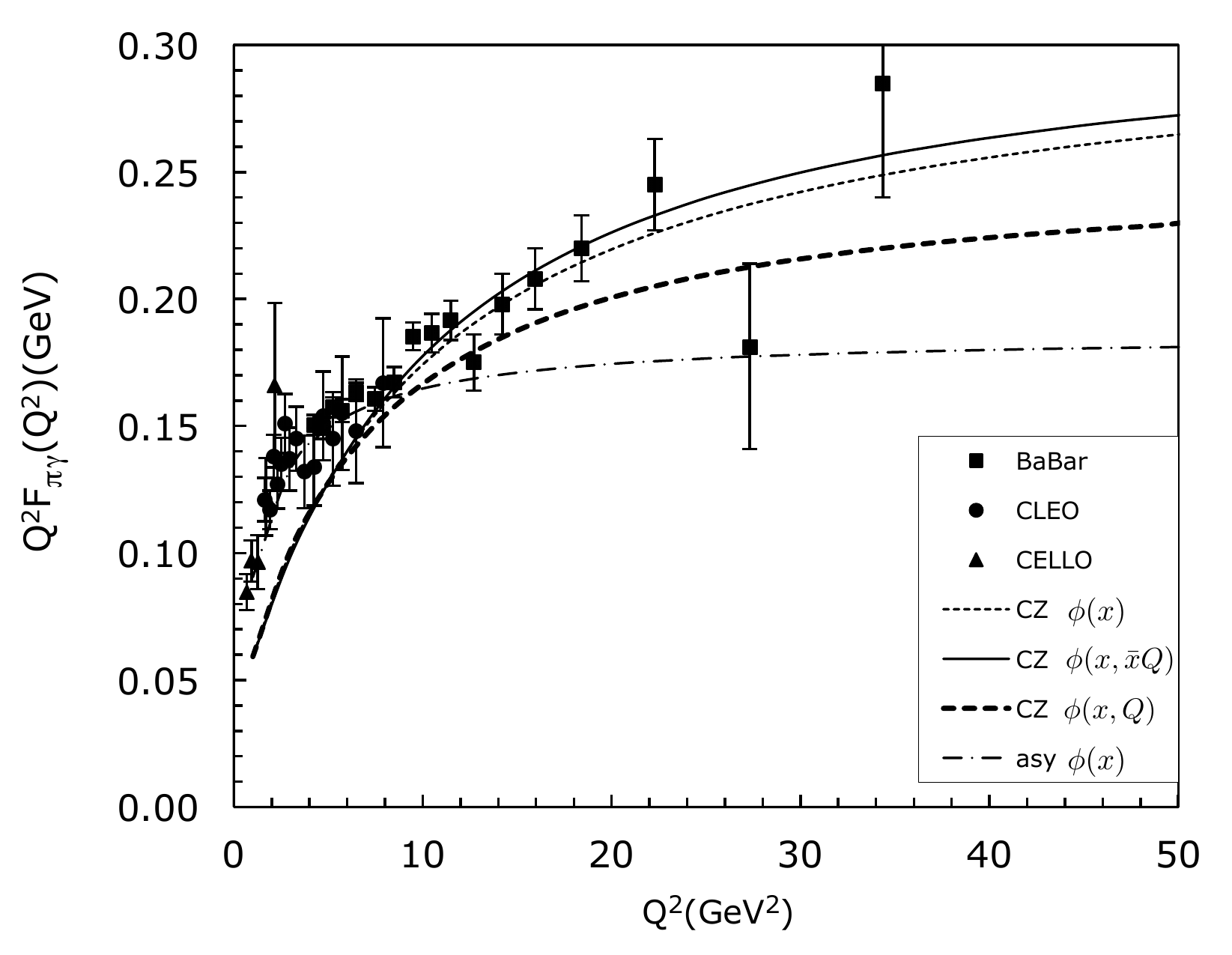}
\caption{Similar as in Fig. \ref{fig:Q2FpiTF_LO_AdS} but for the CZ model for the pion DA.
\label{fig:Q2FpiTF_LO_CZ}}
\end{center}
\end{figure}

We compare results calculated using $\phi(x,\mu_0)$, $\phi(x,{\bar x} Q)$ and $\phi(x,Q)$ in Eq.~(\ref{eq:TFCao2}).
The results for the AdS and CZ models for the pion DA are shown in Figs.~\ref{fig:Q2FpiTF_LO_AdS}
and \ref{fig:Q2FpiTF_LO_CZ}, respectively. 
The valence probability $P_{q \qbar}=0.50$ has been adopted in these calculations.
Using $\phi(x,Q)$ will unjustifiably reduce the predictions substantially for $Q^2>10$ GeV$^2$.
We conclude that the evolution effect at leading order will not bring any large corrections to
the calculation for the pion transition form factors.
It is a good approximation to use $\phi(x,\mu_0)$ in the pQCD calculation for exclusive processes.
This conclusion can be expected to hold when the evolution is considered at NLO as well. 

Analytical expression exists for each term in the sum in Eq.~(\ref{eq:TFCao4}),
though the expression becomes extraordinary long and tedious for large $n$.
The first term corresponds to the results with the asymptotic DA,
\bea
Q^2 F^{AS}_{\pi \gamma}(Q^2)=2 f_\pi \frac{Q^2}{2 \kappa^2} 
\left( 1-  \frac{Q^2}{2 \kappa^2} {\rm e}^{\frac{Q^2}{2 \kappa^2}} \Gamma [0,\frac{Q^2}{2 \kappa^2} ] \right),
\label{eq:TFASCao}
\eea
where $\Gamma[0,x]$ is the incomplete gamma function. At the asymptotic limit $Q^2 \ra \infty$, Eq. (\ref{eq:TFASCao}) 
gives $Q^2 F_{\pi \gamma} (Q^2) \ra 2 f_\pi$ as expected.
A slightly more complicated expression exists for the CZ model for the pion DA.

\subsection{Next-to-leading Order Corrections}

The next-to-leading order corrections have been studied \cite{MelicNP02,AguilaC81,Braaten83,KMR86,MelicMP03}
under the assumption of $\phi(x,\bar xQ) \simeq \phi(x, Q)$, using the standard hard scattering approach when the $k_\perp$-dependence
in the hard scattering amplitude is ignored.
As illustrated in the last section, a properly treatment of evolution is required.
So it is necessary to revisit the NLO calculations with $\phi(x,{\bar x} Q)$.
Assuming that the $k_\bot$-dependence of the LFWF  introduces the same exponential factor for the small $Q$ region\footnote{Dropping
this exponential factor will hardly affect the calculation at large $Q^2$.}
the TFF can be expressed as
\bea
Q^2 F_{\pi \gamma}^{\rm NLO}(Q^2)=\frac{4}{\sqrt{3}} \int_0^1 {\rm d}x \,
T_H(x,Q^2) \, \phi(x,\bar x Q)
\left[ 1-{\rm exp} \left( - \frac{ \bar x Q^2}{2 \kappa^2 x}  \right) \right ],
\label{eq:TFCao5}
\eea
where \cite{MelicNP02,AguilaC81,Braaten83,KMR86,MelicMP03}
\bea 
T_H(x,Q^2)&=&\frac{1}{\bar x}
+\frac{\alpha_s(\mu_R)}{4 \pi}
C_F \frac{1}{\bar x} \left[-9-\frac{\bar x}{x} {\rm ln}{\bar x} + {\rm ln}^2{\bar x}  \right. \nonumber \\
& &
~~~~~~~~~~~~~~~~~~~~~~~\left. +\left( 3+2 {\rm ln} {\bar x} \right) {\rm ln} \left(\frac{Q^2}{\mu^2_R} \right)
\right],
\label{eq:THNLO}
\eea
and $\phi(x,\bar x Q)$ is the `hard' DA evolved at the next-to-leading order \cite{Muller95},
except for the `flat' DA which cannot be evolved as discussed in Section 2.
The regularization scale is commonly taken as $\mu_R=Q$ to eliminate otherwise large logarithm terms.

\begin{figure}		
\begin{center}
\includegraphics[width=120mm]{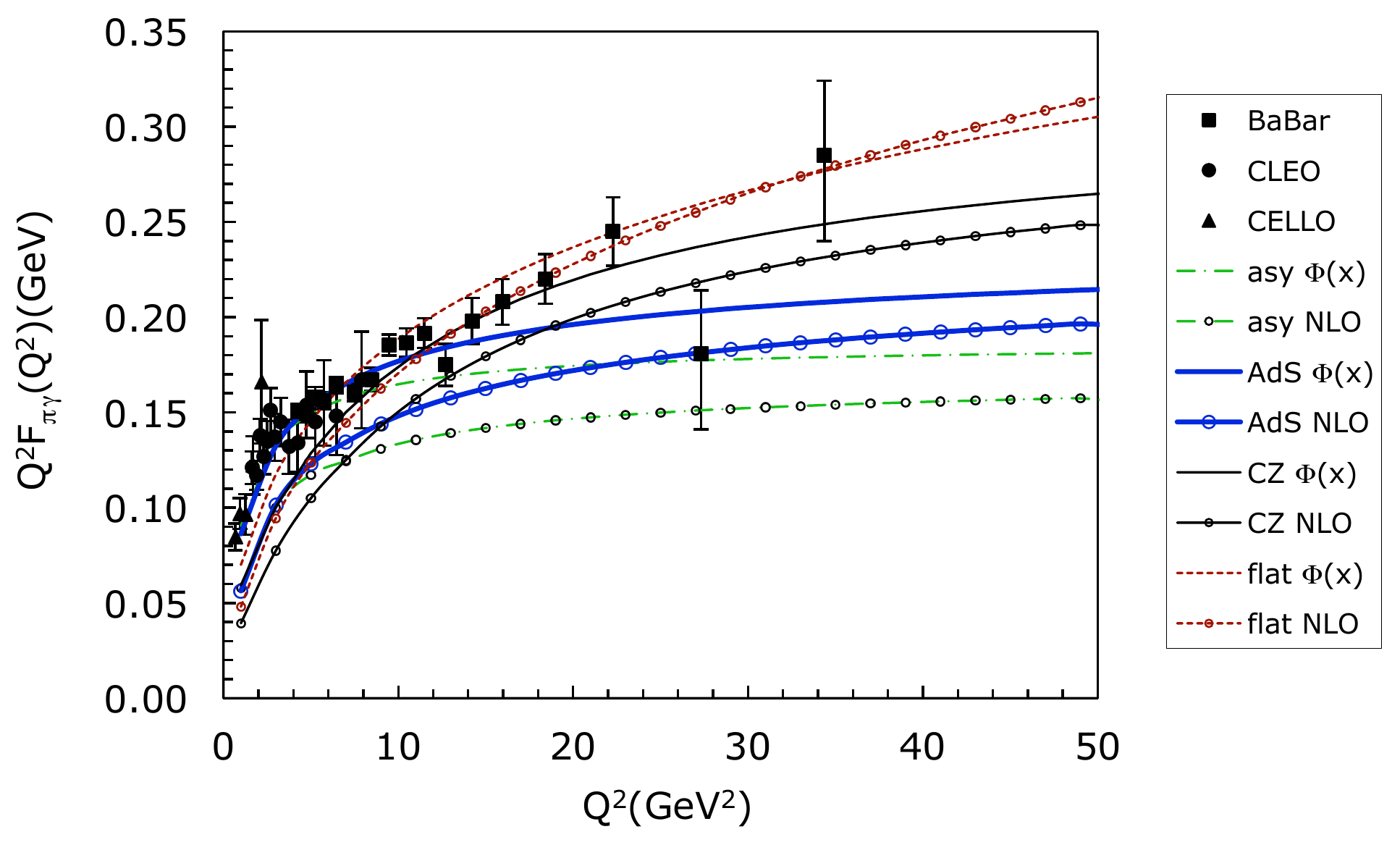}
\caption{
Effect of NLO corrections on the pion to photon transition form factor $Q^2 F_{\pi \gamma}(Q^2)$.
The curves without markers are the results calculated using Eq.~(\ref{eq:TFCao2}) with $\phi(x)$
for the four models of the pion DA:
solid curve -- CZ model,
dashed curve -- `flat' model,
thick-solid curve -- AdS model,
and dash-dotted curve -- asymptotic model.
The curves with markers are the NLO results calculated using Eqs.~(\ref{eq:TFCao5}) and~(\ref{eq:THNLO}).
$P_{q \bar q}=0.50$.
The data are taken from \cite{BaBar_pi0,CELLO,CLEO}.
\label{fig:Q2Fpi_NLO}}
\end{center}
\end{figure}

The numerical results for the pion TFF with the four DA models discussed in Section II  are shown
in Fig.~\ref{fig:Q2Fpi_NLO}.
The NLO corrections vary according to the models used for the pion DA.
The corrections at $Q^2 \sim 5$ GeV$^2$ are about $20\%$, $17\%$, $11\%$, and $15\%$ for the 
asymptotic, AdS, CZ, and `flat'  models. The corrections at $Q^2 \sim 40$ GeV$^2$ are still more
than $7\%$ for all the DAs. Thus it is necessary to take into account these corrections for a large range
of $Q^2$.

\subsection{Higher Order and Higher Fock State Contributions and Dependence on $P_{q\qbar}$}

The calculations for the transition form factors depend on the non-perturbative input, {\it i.e.} the  soft LFWF. Consequently,  it has been long argued that the pion-photon 
transition form factor is a particularly suitable
process in determining the pion LFWF and DA.
As discussed in Section II, we have treated the only parameter $\kappa$ in the model LFWF (Eq.~(\ref{eq:SLFWF})) as a parameter which is constrained,
though not very strictly, by the probability of finding the valence Fock state and the mean square transverse momentum.
In the above calculations we have adopted $P_{q \bar q}=0.50$. The next-to-leading order predictions with $P_{q \bar q}=0.50$
for the pion-photon transition form factor are smaller that the experimental data, particularly for the $Q^2 <10$ GeV$^2$ region.
To improve the agreement between the calculations and experimental data one could use a larger value for $P_{q \bar q}$.
For example, using $P_{q \bar q}=1.0$ will give a much better agreement for the calculations with the CZ model for the pion DA. 
However, a much larger value of $P_{q \bar q}$ than 0.5 will result in a much smaller value for the root of mean square transverse momentum
of the valence quarks compared to the value obtained from the charge radius of the pion.

\begin{figure}	
\begin{center}
\includegraphics[width=90mm]{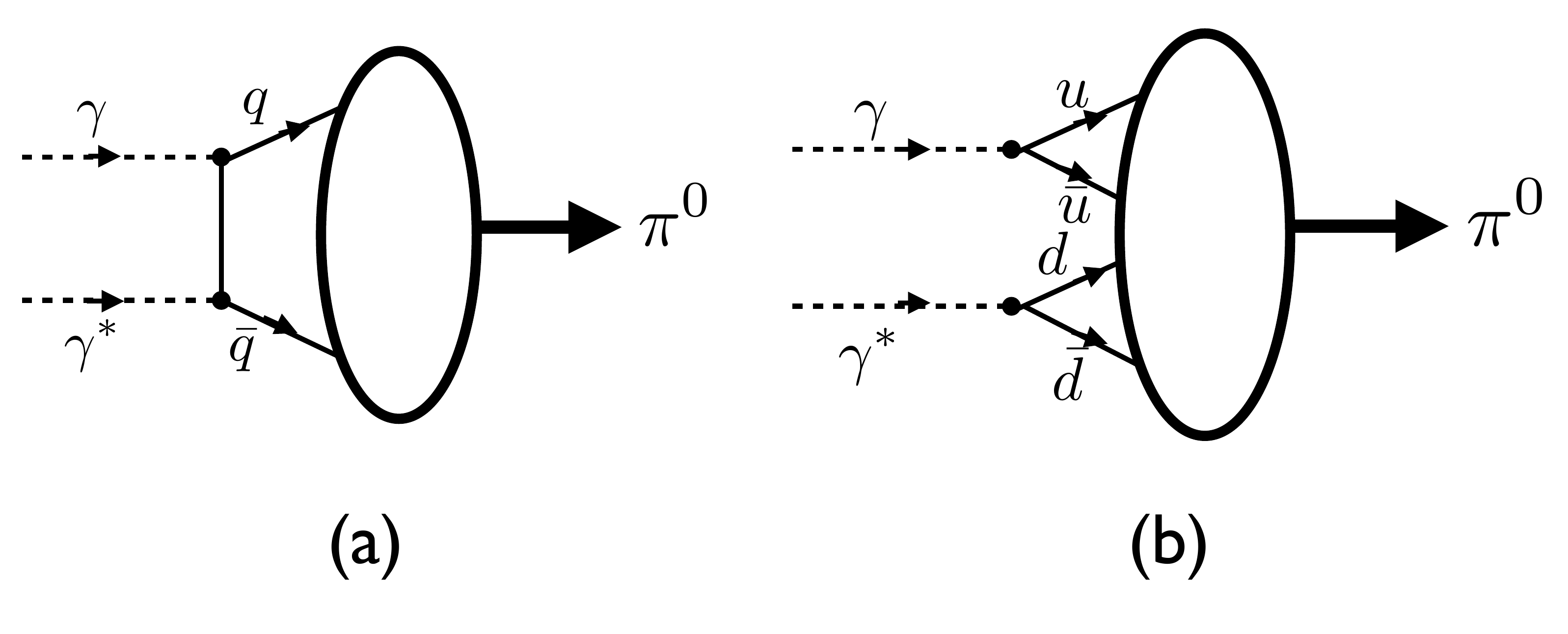}
\caption{Leading-twist contribution (a) and a possible higher-twist contribution (b) to the process $\gamma \gamma^* \ra \pi^0$.
\label{fig:LeadingT+HT}}
\end{center}
\end{figure}

It is shown in \cite{MelicMP03} that the next-to-next-to-leading corrections are much smaller that the next-to-leading corrections.
However, the contributions from higher Fock states ({\it e.g.}, $\vert q \qbar q \qbar \rangle$) are important at low $Q^2$.
Figure \ref{fig:LeadingT+HT} (b) illustrates
such a contribution where 
each photon couples directly to a $q \bar q$ pair.
Such higher-twist contributions $\sum_{e_i \ne e_j} e_i e_j$ are necessary to derive the low energy amplitude for Compton scattering $\gamma H \to \gamma H$, which is proportional to the total charge squared $e_H^2 = (e_i + e_j)^2$ of the target.
These contributions are suppressed by the factor $(1/Q^2)^n$ at large $Q^2$, where $n$ can be understood as the number of $q\qbar$ pairs in the higher
Fock states.
An analysis of these contributions using the framework of AdS/QCD is presented in \cite{BrodskyCTAdS11}.
To estimate these higher Fock state contributions we adopted a phenomenological model as in 
\cite{WuH10}   
\bea
Q^2 F_{\pi \gamma}^{\rm HFS}(Q^2)=\frac{F_{\pi \gamma}(0)/2}{(1+Q^2/\Lambda^2)^2},
\label{eq:Q2FpiHFS}
\eea
where $F_{\pi \gamma}(0)=1/(4\pi^2 f_\pi)$ is the PCAC result and $\Lambda$ can be treated as a parameter.
The contributions are less than $1\%$ for $Q^2>10$ GeV$^2$ and thus can be safely ignored.

\begin{figure}	
\begin{center}
\includegraphics[width=120mm]{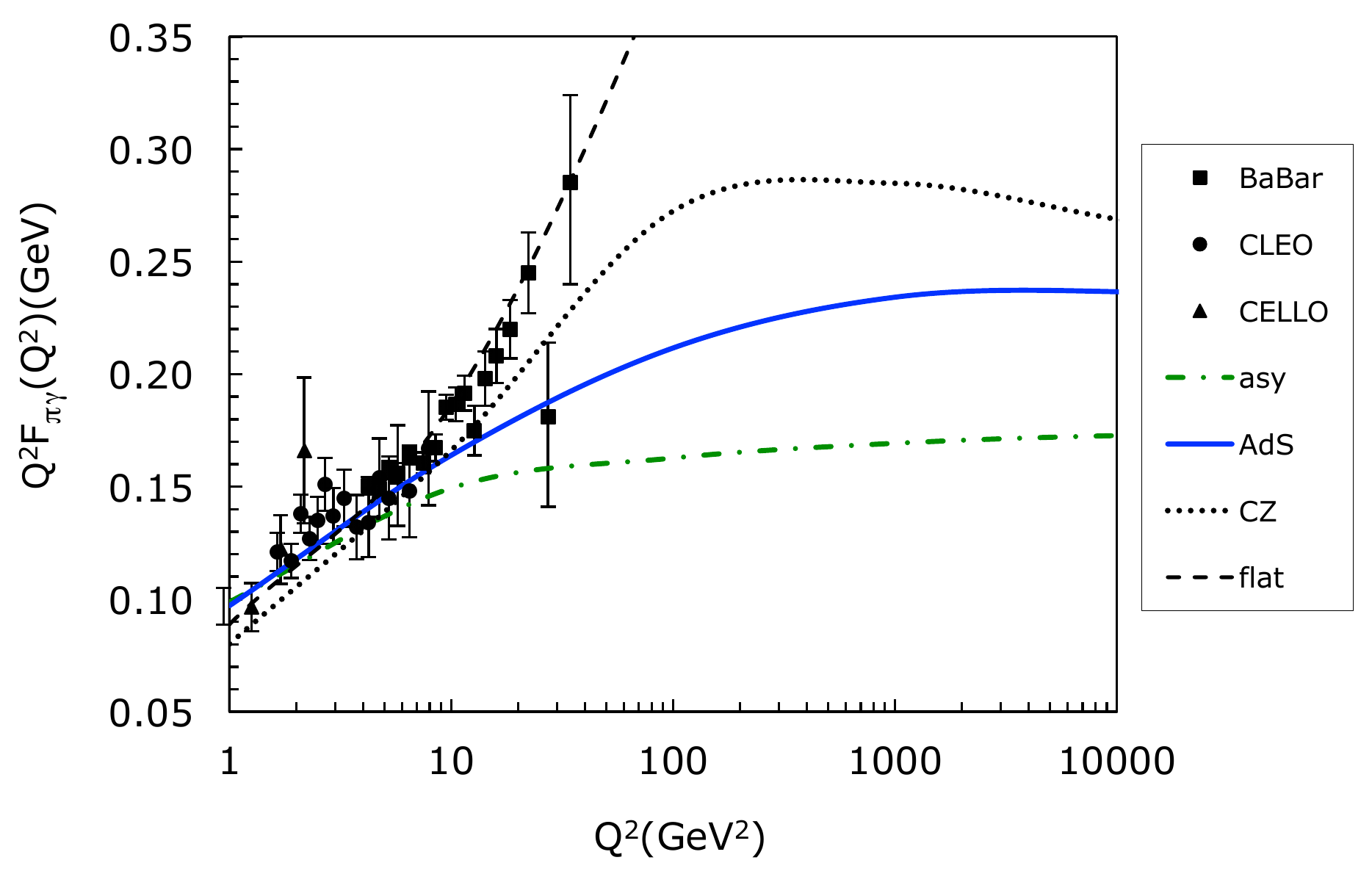}
\caption{The $\pi^0-\gamma$ transition form factor including contributions from the valence Fock state 
(Eq.~(\ref{eq:TFCao5}))
and higher Fock states (Eq.~(\ref{eq:Q2FpiHFS})) of the pion.
The data are taken from \cite{BaBar_pi0,CELLO,CLEO}.
\label{fig:Q2FpiTFF_NLO+HFS}}
\end{center}
\end{figure}

The total contribution from the valence Fock state and the higher Fock states is the sum of  Eqs.~(\ref{eq:TFCao5}) and (\ref{eq:Q2FpiHFS}).
The results calculated with the choice of $P_{q \qbar}=0.5$ and $\Lambda=1.1$ GeV in Eq.~(\ref{eq:Q2FpiHFS}) are compared with the data
in Fig. \ref{fig:Q2FpiTFF_NLO+HFS}.
The agreement at the low $Q^2$ region is vastly improved due to the inclusion of higher Fock state contributions.
However, the higher Fock state contributions are negligible for $Q^2>10$ GeV$^2$ and it is in this large $Q^2$ region that the four models
of the pion DA, discussed in Section II, give very different predictions for the $Q^2$ dependence of the pion-photon transition form factor.
The results with the asymptotic DA are smaller than the \babar\ data and, as expected, do not exhibit a strong
$Q^2$ dependence.
The results with the `flat' DA show a substantial and continuous growth with $Q^2$, which is in disagreement with the QCD prediction
that the pion TFF should approach its asymptotic value of $2 f_\pi$ at $Q^2 \ra \infty$.
In fact,
one cannot apply the ERBL evolution equation to the `flat' DA since it does not satisfy the boundary condition of
the pion DA vanishing at $x=0$ and $x=1$.
The results with the AdS model and CZ model for the pion DA lie in between the predictions of the asymptotic DA and the `flat' DA.
The results with the CZ DA show a fast growth with $Q^2$ 
compared with the AdS DA over the range of $10$ GeV$^2$ $<Q^2< 100$ GeV$^2$.
We note that the \babar\ data for $Q^2>20$ GeV$^2$ suffer larger uncertainties as compared with the low- and medium-$Q^2$ regions.
We also note
that the `flat'  and CZ models of the pion DA will produce much larger values for the $\eta$-photon and $\eta^\prime$-photon transition form factors than
the results reported by the \babar\ Collaboration for $Q^2 > 15$ GeV$^2$ \cite{BaBar_eta,Druzhinin10} and at $Q^2 = 112$ GeV$^2$ \cite{BaBar06}.
Figure~\ref{fig:Q2FpiTFF_NLO+HFS} also shows that the calculations
approach the asymptotic limit value $Q^2 F_{\pi \gamma} (Q^2 \ra \infty)=2 f_\pi$ very slowly since the DA evolution introduces a logarithm $Q^2$-dependence via
$\left[{\rm ln} \left (Q^2/ \Lambda_{\rm QCD}\right)/{\rm ln} \left (\mu_0^2/\Lambda_{\rm QCD}\right )\right]^{-\gamma_n}$.

\begin{figure}	
\begin{center}
\includegraphics[width=120mm]{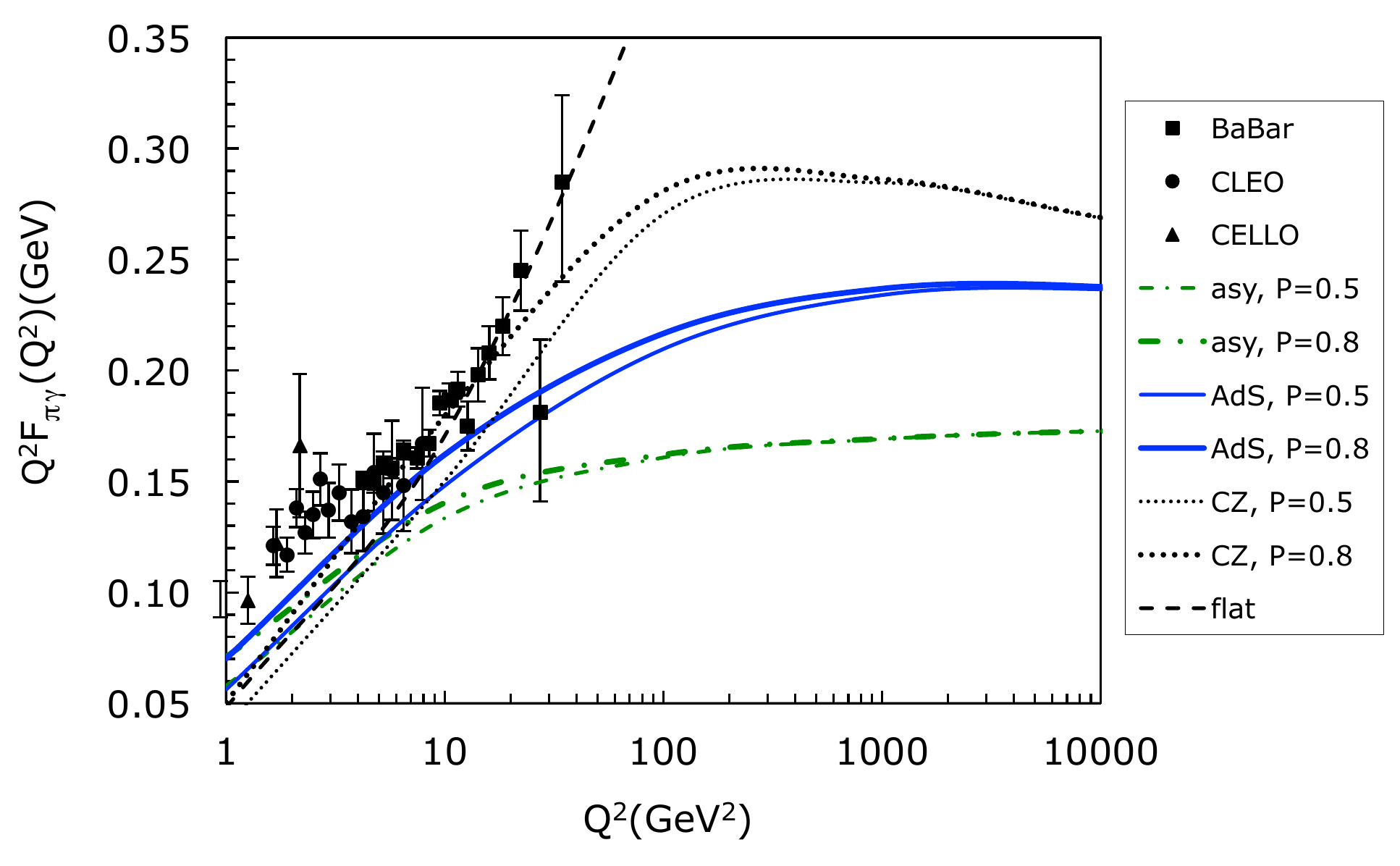}
\caption{The $\pi^0-\gamma$ transition form factor calculated with Eq.~(\ref{eq:TFCao5}) for $P_{q\qbar}=0.5$ and $P_{q\qbar}=0.8$.
The data are taken from \cite{BaBar_pi0,CELLO,CLEO}.
\label{fig:Q2FpiTFF_P50vsP80}}
\end{center}
\end{figure}

We investigate the dependence of our calculations on $P_{q\qbar}$ by allowing $P_{q\qbar}$ to be in the range of $0.5 \sim 0.8$.
The valence Fock state contributions calculated with Eq.~(\ref{eq:TFCao5}) are shown in Fig.~\ref{fig:Q2FpiTFF_P50vsP80}. 
One can see that the calculations for $Q^2>30$ GeV$^2$ depend on $P_{q\qbar}$ very weakly, though
the dependence at the lower $Q^2$ region is much more significant. 
The four models of the pion DA give very different predictions for the pion-photon transition for the region of $Q^2>30$ GeV$^2$, regardless the
value of the $P_{q\qbar}$.

It is very difficult to accommodate the \babar\ large-$Q^2$ data with the QCD calculations using the asymptotic, AdS, and CZ models for the pion DA.
The calculations with the `flat' model of the pion DA can produce a rapid growth for the pion TFF shown by  the \babar\ data. However, the calculations with the same
DA model underestimate significantly the pion TFF at the low $Q^2$, and the prediction for the pion TFF at the asymptotic limit $Q^2 \ra \infty$ 
violates seriously the Brodsky-Lepage limit of $Q^2 F_{\pi \gamma}(Q^2 \ra \infty)=2 f_\pi$.

It was pointed out in \cite{RobertsRBGT10} that using a contact interaction for the quark-antiquark interaction
in the Dyson-Schwinger equations (i.e. treating the pion as a point-like bound state)
produces a `flat' DA and gives predictions for the pion electromagnetic form factors \cite{GGRoberts10} and transition form factor
that are in striking disagreement with completed experiments.
In Ref. \cite{PhamP11} the pion is treated as an elementary field in the triangle graph and the simple expression obtained as
$F_{\pi^0 \gamma}(Q^2) \sim \frac{m^2}{Q^2} ({\rm ln}\frac{Q^2}{m^2})^2$ (with $m=132$ MeV) is able to 
reproduce the \babar\ data for the pion TFF.
We would like to emphasize that although the chiral field theory is a useful approximation for some long-wavelength, soft processes,
it is inapplicable to the hard scattering regime of the \babar\ data. In fact, the compositeness of the pion
in terms of quarks and gluons has been verified in high energy 
experiments both in inclusive reactions (such as the Drell-Yan process for pion-nucleon 
collisions) and many hard exclusive reactions (such as the pion form factor at large spacelike and timelike momentum transfers and large angle scattering processes 
such as $\gamma \gamma \to \pi \pi$ and pion photoproduction).
It is also not necessary to treat the pion as elementary to prove chiral anomalies or the 
Gell-Mann-Oakes-Renner (GMOR) relation.  Such relations are standard consequences of QCD for a composite pion \cite{BrodskyRS10}. 
Employing the BMS models \cite{BMS} for the pion DA, which were determined utilizing the CELLO and CLEO data for the pion TFF,
will produce a $Q^2$-dependence for the pion TFF  similar to that obtained with the AdS model for the pion DA.
A recent analysis \cite{BMPS11} of all existing data (CELLO, CLEO and \babar\ ) performed 
in a framework similar to \cite{BMS} suggest that it is not possible 
to accommodate the high-$Q^2$ tail of the \babar\ data with the same accuracy as the analysis of the CELLO and CLEO data.

We note that there are several theoretical studies 
\cite{WuH10,Kroll10,BroniowshiA10,GorchetinGS11,PhamP11,ADorokhov10,SAgaevBOP11,ZuoJH10,StofferZ11,KotkoP09}
trying to reproduce the \babar\ data for the pion TFF, apart from those using the `flat' form for the pion DA \cite{ADorokhov09,flat,Polyakov09,BaBar_expln_LiM09}.
It was claimed in \cite{WuH10,Kroll10} that a much broader DA than the asymptotic form (but which still vanishes at the end-points) would be able to explain the \babar\ results.
The Regge approach was employed in \cite{BroniowshiA10,GorchetinGS11} to explain the \babar\ data.
On the other hand, there are also theoretical calculations suggesting that the \babar\ data are not compatible with QCD calculations \cite{MikhailovS09,RobertsRBGT10,BMPS11,WuH11}. 

We would like to remark that more accurate measurements of the pion-photon transition form factor at the large $Q^2$ region will be able to distinguish
the various models of the pion DA under discussion.

\subsection{The transition form factor for the pion-virtual-photon}

The above analysis can be easily extended to the case in which the photons involved are both off mass-shell,
{\it i.e.}, for the form factor $F_{\pi \gamma^*}(Q_1^2, Q_2^2)$, by replacing the hard-scattering amplitude $T_H$
with the corresponding expression. 
At leading order $T_H$ has the form
\bea
T_H^{\gamma^* \gamma^* \ra \pi^0}(x,Q_1,Q_2)=\frac{1}{ \bar x Q_1^2 + x Q_2^2},
\label{eq:THvirtual}
\eea
where $\bar{x}=1-x$. For the expression at next-to-leading order we refer the
readers to reference~\cite{Braaten83}.

A significant difference from the case with a real and a virtual photons where $T_H=1/(\bar x Q^2)$ is that
Eq.~(\ref{eq:THvirtual}) is not divergent at the end-points. 
Thus considering the $k_\perp$-dependence will not bring as large corrections as for $F_{\pi \gamma}(Q^2)$,
and  the transition form factor $F_{\pi \gamma^*}$ is much less sensitive to the
end-point behavior of the pion DA than $F_{\pi \gamma}$.

We note that the kinematic region satisfying $Q_1^2=Q_2^2$ is 
particularly interesting since in this region the 
amplitude $T_H$ becomes
independent of $x$ and thereby the transition form factor is largely
described by the normalization of the pion DA, which is model and
$Q^2$ independent.
Ignoring the weak $Q^2$ dependence introduced by the consideration of $k_\perp$ dependence and the NLO corrections in $\alpha_s$,
which are both expected to be small at large $Q^2$, we have
\bea
Q_1^2 F_{\pi \gamma^*}(Q_1^2,Q_2^2) \ra \frac{2}{3} f_\pi \,\, \makebox{for $Q_1=Q_2 >$ a few GeV}.
\label{eq:FpivirtualCao}
\eea
We make the remark that Eq.~(\ref{eq:FpivirtualCao}) is expected to work 
for the range $Q_1^2=Q_2^2 \sim$ 10-20 GeV$^2$ which is accessible by the current experiments.
So measurements of $F_{\pi \gamma^*}(Q_1^2,Q_2^2)$ under these conditions would provide another test
of pQCD analysis of exclusive processes.

\begin{figure}		
\begin{center}
\includegraphics[width=100mm]{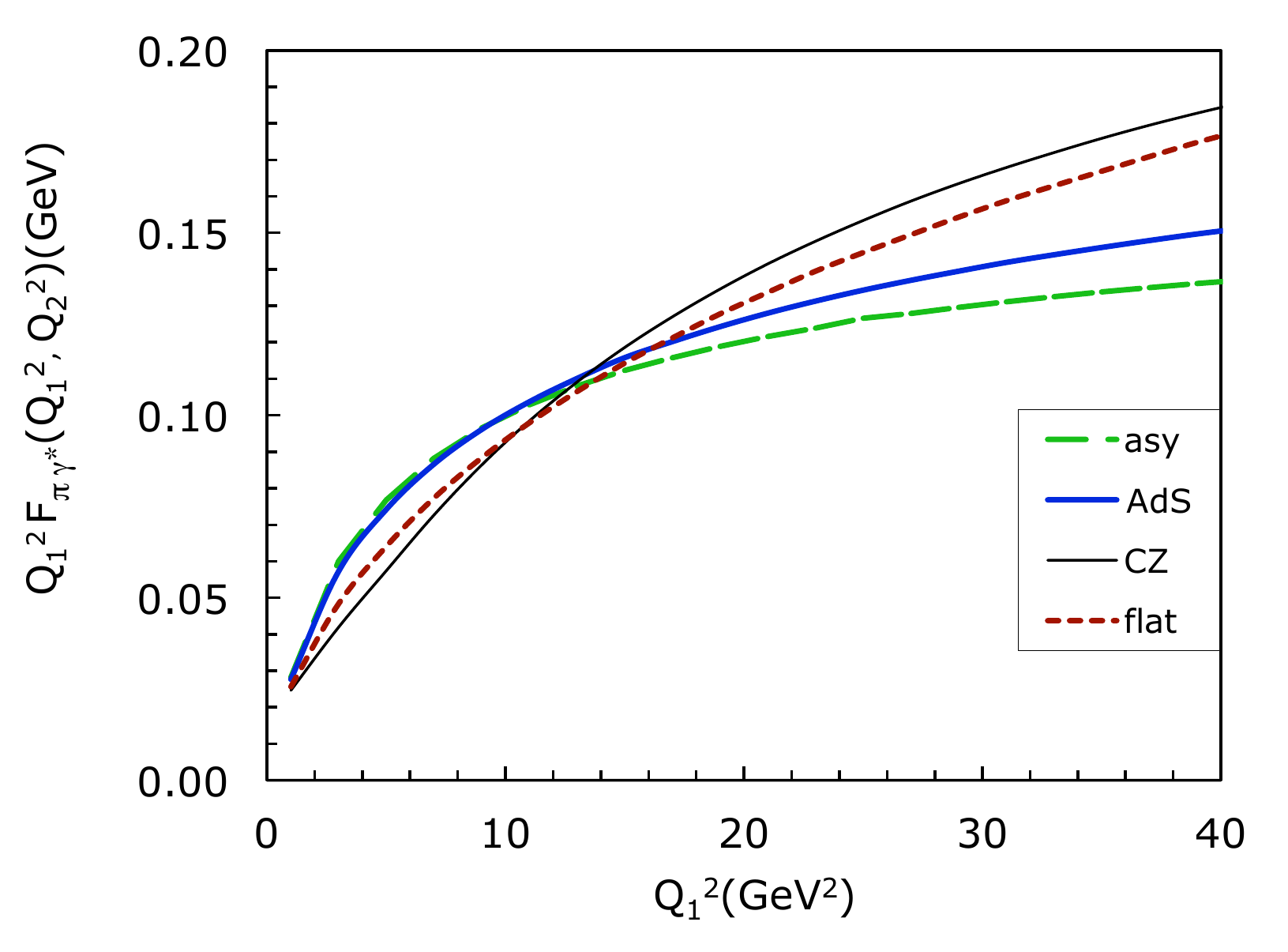}
\caption{
The doubly virtual transition form factor  for $Q_2^2=2$ GeV$^2$, calculated with Eqs.~(\ref{eq:TFCao5}) and (\ref{eq:THvirtual}).
The solid, dashed, thick-solid, and long-dashed curves are the results with
the CZ, `flat', AdS, and asymptotic models for the pion DA. }
\label{fig:Q2Fpivirtual}
\end{center}
\end{figure}

The numerical results for $F_{\pi \gamma^*}(Q_1^2,Q_2^2)$ calculated at NLO are given in Fig.~\ref{fig:Q2Fpivirtual}
for $Q_2^2=2$ GeV$^2$. 
The four models give similar predictions for $Q_1^2$ up to $15$ GeV$^2$.
At large $Q_1^2$ the results with the CZ model are much larger
compared with the asymptotic and AdS models.
We found that the NLO corrections at this range of $Q^2$ are less than $10\%$ for the asymptotic and AdS models while the corrections
to the CZ and `flat' models are negligible. The higher-twist effects at this range of $Q^2$ could be expected to be minimal.
Thus the difference on the prediction for this transition form factor is a direct reflection of different behavior of the pion DA.
Measurements of this form factor at the kinematic region $Q_1^2 \sim 20~Q_2^2$ with $Q_2^2$ being about 
a few GeV$^2$ would provide a good laboratory to distinguish the middle-peak DA, such as the asymptotic and AdS models,
from the CZ model.

\section{The $\eta$-photon and $\eta^\prime$-photon transition form factors}

According to the SU(3)$_F$ quark model, the three charge neutral states in the nonet of pseudoscalar mesons are 
$\pi^0$, $\eta_8$ and $\eta_1$. The latter two mix to give the physical particles $\eta$ and $\eta^\prime$.
It can be expected that the states $\pi^0$, $\eta_8$ and $\eta_1$ have the same form of distribution amplitude (and the same
$k_\perp$-dependence in the light-front wavefunctions), 
\bea
\phi_P(x)=f_P f(x),
\eea
with $P$ denoting $\pi^0$, $\eta_8$ and $\eta_1$, and $f_P$ being the corresponding decay constant.

The transition from factors for the $\pi^0$, $\eta_8$ and $\eta_1$ can be expressed as
\bea
Q^2 F_{P \gamma}(Q^2)=\frac{4}{\sqrt{3}} c_P \int_0^1 {\rm d}x \, T_H(x,Q^2) \phi_P(x, {\bar x} Q)
\left[ 1-{\rm exp} \left( - \frac{ \bar x  Q^2}{2 \kappa^2 x }  \right) \right ],
\label{eq:Q2TFF}
\eea
where $c_P=1, \, \frac{1}{\sqrt{3}}$, and $\frac{2\sqrt{2}}{\sqrt{3}}$ for $\pi^0$, $\eta_8$ and $\eta_1$, respectively
and $T_H(x, Q^2)$ is given by Eq.~(\ref{eq:THNLO}) at next-leading order of QCD running coupling constant.

\begin{figure}	
\begin{center}
\includegraphics[width=100mm]{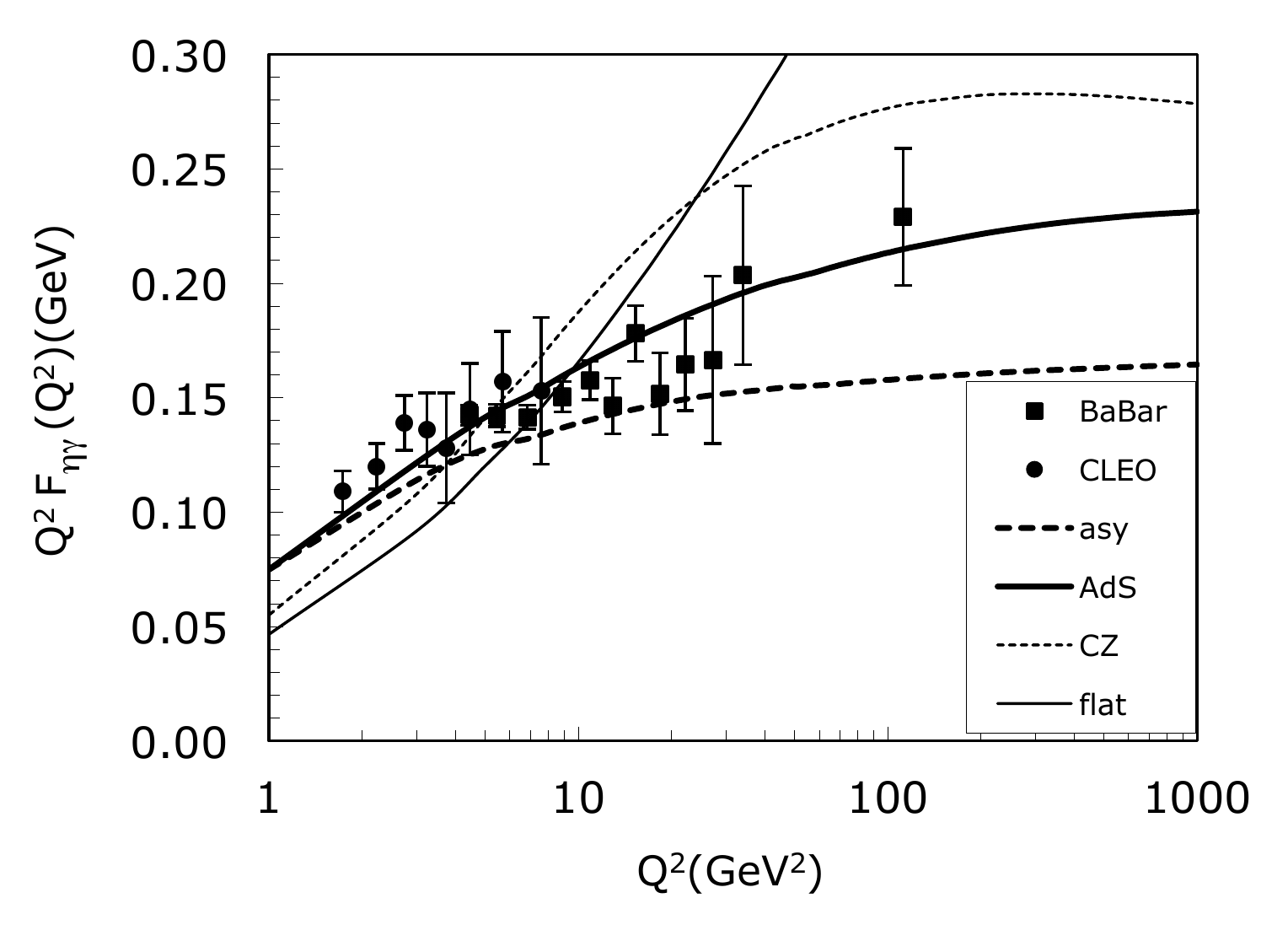}
\caption{The $\eta - \gamma$ transition form factor shown as $Q^2 F_{\eta \gamma}(Q_2)$.
The thick-dashed, thick-solid, thin-dashed, thin-solid curves are the results calculated with the
asymptotic, AdS, CZ and `flat' models for the meson DAs, respectively.
Data are taken from \cite{CLEO} (CLEO) and \cite{BaBar_eta,Druzhinin10} (\babar).\label{fig:Q2Feta}}
\end{center}
\end{figure}

\begin{figure}	
\begin{center}
\includegraphics[width=100mm]{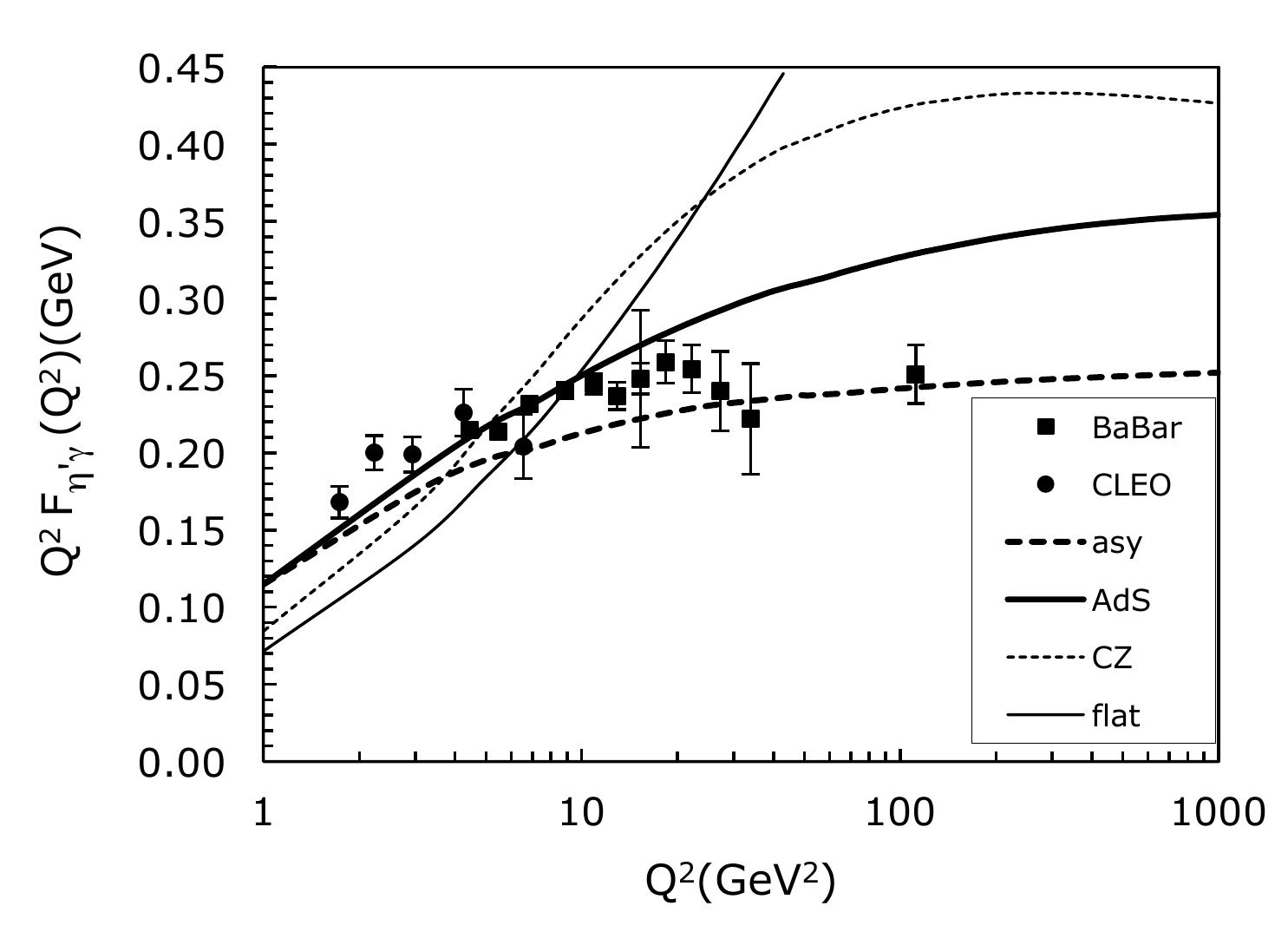}
\caption{The $\eta^\prime-\gamma$ transition form factor shown as $Q^2 F_{\eta^\prime \gamma}(Q_2)$.
The thick-dashed, thick-solid, thin-dashed, thin-solid curves are the results calculated with the
asymptotic, AdS, CZ and `flat' models for the meson DAs, respectively. 
Data are taken from \cite{CLEO} (CLEO) and \cite{BaBar_eta,Druzhinin10} (\babar).\label{fig:Q2FetaP}}
\end{center}
\end{figure}

The transition form factors for the $\eta$ and $\eta^\prime$ result from the mixing of $F_{\eta_8 \gamma}$ and $F_{\eta_1 \gamma}$,
\bea
\left(
	\begin{array}{c}
	F_{\eta \gamma} \\
	F_{\eta^\prime \gamma}
	\end{array}
\right)
=\left(
	\begin{array}{cc}
	{\rm cos} \, \theta & -{\rm sin} \, \theta \\
	{\rm sin} \, \theta & {\rm cos} \, \theta
	\end{array}
\right)
\left(
	\begin{array}{c}
	F_{\eta_8 \gamma} \\
	F_{\eta_1 \gamma}
	\end{array}
\right),
\eea
where $\theta$ is the mixing angle which has been the subject of extensive studies \cite{theta_review}.
In this work we adopt $\theta=-14.5^o\pm 2^o$, $f_8=(0.94 \pm 0.07) f_\pi$, and $f_1=(1.17 \pm)  f_\pi$ \cite{CaoSignal99}.
The same value of $\kappa$ has been used for the three charge neutral states and $\Lambda=1.1$ GeV is adopted in Eq.~(\ref{eq:Q2FpiHFS}).
The results for the $\eta$-photon and $\eta^\prime$-photon transitions form factors are shown in
Figs. \ref{fig:Q2Feta} and \ref{fig:Q2FetaP} respectively.
The data favor the AdS and the asymptotic models for the meson DA.
One may fine-tune the parameters $\theta$, $f_8$, and $f_1$ to make the calculations with the asymptotic form and the AdS form
to give better agreement with the data. However it is almost impossible to make the calculations with the CZ form
and the `flat' form to describe the data in both the low- and high-$Q^2$ regions simultaneously for the two transition form factors, although
these two forms are favored to explain the rapid growth of the \babar\ data \cite{BaBar_pi0} for the pion-photon transition form factor at
large $Q^2$. 

The DA models that could explain the \babar\ measurements for the pion-photon transition form factors at large values of $Q^2$
will fail in QCD calculations for the other processes, including the $\eta$-photon and $\eta^\prime$-photon transition form factors
reported by the \babar\ Collaboration.
This may suggest that the \babar\ measurements at large $Q^2$ are not a true accurate representation of
the pion-photon transition form factor, a perspective that has been suggested in \cite{MikhailovS09,RobertsRBGT10}.
This may also indicate that there are some inconsistencies among the results for the transition form factors of the $\pi^0$, $\eta$ and $\eta^\prime$.

 In a recent paper\cite{WuH11}, Wu and Huang have studied the dependence of the photon-to-meson transition form factors
 on the model parameters (including quark masses, mixing angle, as well as an intrinsic charm component) in a light-front perturbative approach.
 It is found that the agreement of the predictions of their model with the experimental data can be somewhat improved by adjusting these parameters within their reasonable regimes, but
 the data for the pion-photon transition form factor in the entire $Q^2$ domain cannot be explained consistently.

Recently Kroll \cite{Kroll10} analyzed the $\pi-\gamma$, $\eta-\gamma$, and $\eta^\prime-\gamma$ transition form factors
using the modified hard scattering approach in which the transverse-momentum-factorization is combined with a Sudakov factor. 
The distribution amplitude of mesons is constrained to contain the
first three nontrivial terms in the Gegenbauer expansion for the DA, and
a Gaussian form is assumed for the $k_\perp$-dependence in the wave function. By adjusting the three parameters --
the two coefficients, $a_2$ and $a_4$ in the Gegenbauer expansion, and the transverse size parameter $\sigma_P$,
reasonably good agreement with experimental data was achieved.
The best fit
for the pion-photon transition form factor presented in Fig. 4 of \cite{Kroll10}
is very similar to our results calculated with the AdS model for the pion DA (see Fig.~\ref{fig:Q2FpiTFF_NLO+HFS}).
We note that in order to describe the three transition form factors, very different values for the three parameters are chosen
in  Ref.~\cite{Kroll10} for the $\pi_0$, $\eta_8$ and $\eta_1$:
$a_2^\pi=0.20$, $a_4^\pi=0.01$, $\sigma^\pi=0.40$ GeV$^{-1}$;
$a_2^8=-0.06$, $a_4^8=0$, $\sigma^8=0.84$ GeV$^{-1}$;
$a_2^1=-0.07$, $a_4^1=0$, and $\sigma^1=0.74$ GeV$^{-1}$.
Such a choice of parameters suggests a very large SU(3)$_F$ symmetry breaking between the DAs of the $\pi^0$ and $\eta^8$
and a very little SU(3)$_F$ symmetry breaking between the $\eta_8$ and $\eta_1$. 
Our view is that this remains as a possibility, but it is unlikely since the $\pi_0$ and $\eta_8$ belong to the octet and $\eta_1$ belongs to the singlet of the pseudoscalar mesons.
Furthermore, there is no evidence supporting such a large SU(3)$_F$ breaking in other processes,
 {\it e.g.}, decay processes involving pseudoscalar mesons \cite{theta_review}.
In fact, the CLEO's measurements of the meson-photon transition form factors \cite{CLEO} suggested that the $\pi^0$ and $\eta_8$  
have very similar non-perturbative dynamics and thereby similar light-front wavefunctions and distribution amplitudes.

\section{Summary}

The photon-to-meson transition  form factor measured in $\gamma^*  \gamma \to M$ is the simplest hadronic amplitude predicted by QCD. 
Measurements from electron-positron colliders provide  important constraints on the non-perturbative hadron distribution amplitude, a fundamental gauge-invariant measure of hadron structure.
The meson distribution amplitude $\phi(x,Q)$ evolves in pQCD according to the ERBL evolution equation, which is based on first-principle properties of QCD.
Important constraints on the distribution amplitude have been obtained using lattice gauge theory.
We have analyzed in detail four models for the $\pi^0, \eta$, and  $\eta^\prime$ distribution amplitudes that have been suggested in the literature,
including their QCD evolution with $\log Q^2$. 

We have calculated the meson-photon transition form factors for the $\pi^0$, $\eta$ and $\eta^\prime$,
taking into account effects which are important for the calculations at finite $Q^2$. These effects include the $k_\bot$-dependence
of the hard-scattering amplitude and light-front wavefunctions, the evolution effects of the pion distribution amplitude,
and NLO corrections in $\alpha_s$. We have pointed out that a widely-used approximation of replacing $\phi(x, {\bar x} Q)$ with $\phi(x,Q)$ in the hard-scattering
formalism will significantly, and unjustifiably, reduce the predictions for the magnitude of hard exclusive amplitudes.

It is found that in order to explain the experimental data at $Q^2<10$ GeV$^2$ one needs to take into account the contributions from higher Fock states of the mesons, although 
these contribution are negligible for the larger $Q^2$ region.
The four models of the meson DA discussed in this article
give very different predictions for the $Q^2$ dependence of the meson-photon transition form factors in the large $Q^2$ region.
The predictions based on the AdS/QCD and light-front holography for the pion distribution amplitude agree well
with the experimental data for the $\eta$- and $\eta^\prime$-photon transition form factors, but they disagree 
with the data for the pion-photon transition form factor reported by the \babar\ Collaboration.
The calculations with the CZ model agree with the \babar\ data for the pion-photon transition form factor reasonably well, but 
the predictions  are much larger than the data from the CLEO and \babar\ Collaborations for the $\eta$- and $\eta^\prime$-photon transition form factors.
The calculations with the `flat'  distribution amplitude, which has been advocated in explaining the \babar\  large-$Q^2$ data for the pion transition form factor,
disagree strongly with the CLEO and \babar\ data for the $\eta$- and $\eta^\prime$-photon transition form factors.

We investigated the dependence of the calculations on the probability of valence Fock state of the pion $P_{q\qbar}$. It was found that
the four models of the meson DA give very different predictions for the meson-photon transition form factor in the region of $Q^2 > 30$ GeV$^2$ for  
$P_{q\qbar}$ to be in the reasonable range of $0.5\sim 0.8$.
More accurate measurements of the meson-photon transition form factor in the large $Q^2$ region will be able to distinguish
the four commonly used models of the pion DA.

The \babar\ data for the pion-photon transition from factor exhibit a rapid growth at high $Q^2$, but this feature is 
missing for the $\eta$- and $\eta^\prime$-photon transition form factors.  
The rapid growth of the large-$Q^2$ data for the pion-photon transition form factor reported by the \babar\ Collaboration is 
difficult to explain within the current framework of QCD. This is a viewpoint first expressed by Roberts {\it et al}. \cite{RobertsRBGT10}
in their  Bethe-Salpeter/Dyson-Schwinger analysis of  the pion-photon transition form factors.
If the \babar\ data for the meson-photon transition form factor for the $\pi^0$ is confirmed, 
it could indicate physics beyond-the-standard model,
such as a weakly-coupled elementary  $C=+$ axial vector or pseudoscalar $z^0$ in the few GeV domain, an elementary field which would provide the coupling
$\gamma^* \gamma \to z^0 \to \pi^0$ at leading twist.
We would like to remark that a high-mass state of about $10$ GeV has been envisaged in \cite{BroniowshiA10} to explain the \babar\ data for the pion TFF  \cite{Refereenote}.
We thus emphasize the importance of additional measurements of the meson-photon transition form factors.
 
\begin{acknowledgments}
F. G. Cao is grateful to X.-H. Guo at Beijing Normal University and H. Chen at Southwest University, China for their hospitality
where part of F. G. Cao's work was done.
We thank  C. D. Roberts, N. Stefanis, and V. Braun for helpful comments.
This research was supported by the Department of Energy contract
DE--AC02--76SF00515.   \end{acknowledgments}

\end{document}